\documentclass[Journal]{IEEEtran}
\usepackage{color,graphicx}
\usepackage{url}
\usepackage{soul}
\usepackage{dblfloatfix}
\usepackage{amsmath}
\usepackage{algorithmic}
\usepackage{algorithm}
\usepackage{epstopdf}
\usepackage{epsfig}
\usepackage{cite}
\usepackage{amsmath}
\usepackage{algorithmic}
\usepackage{algorithm}
\usepackage{epstopdf}
\usepackage{epsfig}
\usepackage{cite}

\usepackage{mathtools}
\usepackage{breqn}
\usepackage{amsfonts}
\usepackage{array}
\usepackage{caption}
\usepackage{subcaption}
\usepackage{cuted}
\definecolor{Gray}{gray}{0.9}
\usepackage[table]{xcolor}

\ifCLASSINFOpdf
\else
\fi

\makeatletter

\newcommand{\Rmnum}[1]{\expandafter\@slowromancap\romannumeral #1@}
\makeatother
\begin{document}
\title{Modeling and Performance Analysis of Spatially
	Distributed LTE-U and Wi-Fi Networks}

\author{\IEEEauthorblockN{Anand M. Baswade$^*$, Mohith Reddy$^\dag$,   Antony Franklin A$^*$, and Bheemarjuna Reddy Tamma$^*$}\\
	\IEEEauthorblockA{ $^*$Indian Institute of Technology Hyderabad, India.\\ $^\dag$Indian Institute of Technology BHU, India.\\
		Email: [cs14resch11002$^*$, tbr$^*$, antony.franklin$^*$]@iith.ac.in}, mbmohith.reddy.eee16$^\dag$@itbhu.ac.in }

\maketitle
\begin{abstract}
To access an unlicensed channel \mbox{Wi-Fi} follows Listen Before Talk (LBT) mechanism whereas \mbox{LTE-U} adopts ON-OFF duty cycled mechanism to fairly share the channel with \mbox{Wi-Fi}. These contrasting mechanisms result in quite different performance for \mbox{Wi-Fi} and \mbox{LTE-U} based on their relative deployment and density in the environment. In this work, we present an analytical model for characterization of achievable throughputs of \mbox{Wi-Fi} and \mbox{LTE-U} networks in spatially distributed high-density scenarios. The proposed model is used to study how \mbox{LTE-U} and \mbox{Wi-Fi} coexist with each other in different deployment scenarios. Our extensive simulation results prove it to be a reliable model for estimating throughput of both Wi-Fi and \mbox{LTE-U}. We record a very good accuracy in throughput estimation and the mean normalized error is less than 1\% for 40-node scenario in which 50\% of nodes belong to each of Wi-Fi and \mbox{LTE-U}. Finally, we use the analytical model to conduct coexistence studies of LTE-U and Wi-Fi.

\end{abstract}

\begin{IEEEkeywords}
	\mbox{Wi-Fi}, \mbox{LTE-U}, CSAT, LBT, Inter-RAT Coexistence, 5G, Performance analysis.
\end{IEEEkeywords}

\section{Introduction}\label{Sec1}
The demand for mobile data is growing at a very rapid rate, and is expected to cross 77.5~EB per month by 2022~\cite{Cisco}. Meeting such a huge data demand is a very challenging task for mobile operators due to the high cost of licensed spectrum and other reasons. One of the promising solutions is the usage of unlicensed spectrum for operation of LTE and upcoming 5G Radio Access Technologies (RATs). The major challenge for deploying \mbox{LTE} in unlicensed spectrum is the fair coexistence requirements with existing technologies in the unlicensed spectrum, mainly IEEE 802.11 a.k.a. \mbox{Wi-Fi} technology. Thus, fairness in unlicensed spectrum needs to be properly defined and tested before deploying LTE in unlicensed spectrum. The 3rd Generation Partnership Project (3GPP)~\cite{3GPP} defines fairness as: \textit{LTE design in unlicensed spectrum should be in such a way that it should not impact Wi-Fi more than another Wi-Fi on the same unlicensed channel.} 
\par There are two ways of operating LTE in unlicensed spectrum: (i) LTE with discontinuous transmissions (ON-OFF cycles) a.k.a. LTE-U\cite{8,N1}; (ii) LTE with Listen-Before-Talk (LBT) mechanism a.k.a. Licensed Assisted Access (LAA)~\cite{3GPP} which is similar to Wi-Fi in terms of channel access.
\mbox{LTE-U} has received a lot of attention in the industry because of its ease of implementation. In the ON-OFF cycle of LTE-U, the ON time of \mbox{LTE-U} corresponds to \mbox{LTE-U} transmissions period, and OFF time corresponds to no transmissions from \mbox{LTE-U} so that \mbox{Wi-Fi} can get access to the medium for transmissions. Due to the regulatory restrictions on transmission power in unlicensed spectrum, mobile operators have to deploy a large number of \mbox{LTE-U} small cells in indoor/outdoor environments. Such a dense deployment of LTE-U small cells along with existing Wi-Fi Access Points (APs) could lead to inefficient utilization of unlicensed spectrum resources, due to interference, inter-RAT hidden terminal problem, and lack of coordination among heterogeneous~RATs~\cite{BI}.

\par Most of the existing works on throughput estimation of \mbox{Wi-Fi} assumes that every node in the Wi-Fi network can sense all the other nodes in the network~\cite{bianchi}. Further, to calculate the throughput of \mbox{Wi-Fi} network in a generalized scenario, new models have been proposed~\cite{boe} to cover all possible scenarios \emph{e.g.,} a node can only sense transmissions of a subset of nodes. The recent inclusion of \mbox{LTE-U} in unlicensed spectrum makes throughput estimation of Wi-Fi more challenging. In~\cite{WL} and~\cite{info}, the authors modeled throughputs of \mbox{LTE-U} and \mbox{Wi-Fi} networks in the coexistence scenario where each node can sense all the other nodes on the channel (\emph{i.e.,} both the \mbox{LTE-U} and \mbox{Wi-Fi} nodes are inside Energy Detection Threshold (EDT) range of each other). 
But, the case when the nodes can sense the presence of only a subset of nodes in the network makes throughput estimation more challenging. In practical deployment scenarios, all LTE-U and Wi-Fi nodes cannot hear each other as the nodes are spatially distributed. Hence, throughput estimation in spatially distributed scenarios is essential as it will help to study the coexistence of LTE-U and Wi-Fi in greater depth.
\par In this paper, we model the throughput of both \mbox{LTE-U} and \mbox{Wi-Fi} nodes (eNB or AP) in a spatially distributed scenario. Further, we use the model to study how LTE-U and Wi-Fi coexist with each other in different deployment scenarios. To the best of our knowledge this is the first work in the direction of estimation of throughputs in Wi-Fi--LTE-U networks in spatially distributed scenarios.
\begin{figure*}[b]
	\begin{center}
		\includegraphics[width=16cm,height=4.5cm]{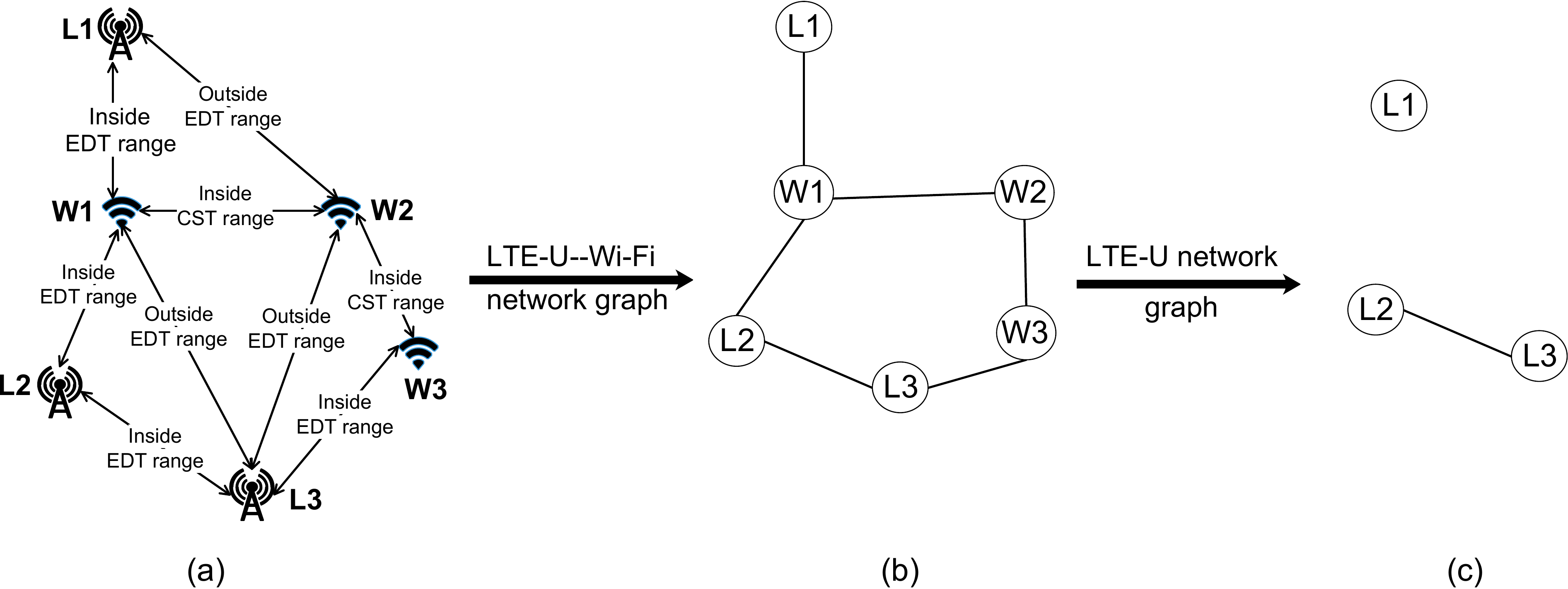}
		\newline
		\caption{(a) An example of spatially distributed scenario consisting of LTE-U and Wi-Fi nodes, (b) its associated LTE-U--Wi-Fi network graph, and (c) its associated LTE-U network~graph.}
		\label{ExampleNet}
	\end{center}
\end{figure*}
The major contributions of the paper are given below:
\begin{enumerate}
	\item We provide an analytical model to evaluate the throughputs of spatially distributed dense \mbox{LTE-U} and \mbox{Wi-Fi} networks.

	\item We validate the proposed analytical model through extensive simulation studies.
	\item Using our analytical model, we perform coexistence studies of \mbox{LTE-U} and Wi-Fi in different deployment scenarios.

\end{enumerate}
\par The rest of the paper is organized as follows. Related works are presented in Section~\ref{RW} and system model is given in Section~\ref{SM}. Section~\ref{Motivation} formulates the problem of throughput estimation in a spatially distributed LTE-U and Wi-Fi network scenario. Section~\ref{model} proposes an analytical model for the same. Section~\ref{PE} validates the proposed model and studies the performance of Wi-Fi--LTE-U networks in different scenarios. Finally, Section~\ref{conclusion} summarizes and concludes the~work.
\section{Related Work}\label{RW}
The authors in~\cite{sagari12} have demonstrated that the effect on Wi-Fi performance is much more when LTE operates in unlicensed without employing any coexistence mechanism. Whereas LTE gets less affected due to the presence of \mbox{Wi-Fi} networks. In fact, it is shown that when both operate on the same channel, Wi-Fi throughput suffers from 20\% to 97\% while LTE experiences loss in throughput up to 10\%. Similar studies on performance degradation to Wi-Fi in the presence of LTE are carried out in~\cite{LWS11},~\cite{LWS22},~\cite{3},~\cite{E1}. 

Thus, for better coexistence with \mbox{Wi-Fi}, duty cycled \mbox{LTE-U} and LBT based LAA are proposed by LTE-U Forum~\cite{N1} and 3GPP~\cite{3GPP}, respectively. Duty cycled \mbox{LTE-U} solution is easy to implement and required very minimal changes in LTE protocol stack hence gained a lot of attention from the industry. An almost blank subframe is one way to mute LTE-U transmission\cite{blank}. Many studies~\cite{N1,E2,E3} demonstrated that the duty cycle mechanism allows LTE to be a fair neighbor to \mbox{\mbox{Wi-Fi}} if the \mbox{LTE-U} ON and \mbox{LTE-U} OFF periods are chosen properly based on the \mbox{Wi-Fi} activities in the unlicensed channel\cite{E3}. 
\par In the literature, there has been a lot of work on modeling and performance analysis of \mbox{Wi-Fi} networks~\cite{bianchi},\cite{boe},\cite{W2},\cite{W3}. The coexistence of LAA and Wi-Fi is modeled in~\cite{A1},~\cite{A4},~\cite{A3} based on widely used Bianchi model\cite{bianchi}. Further, the coexistence of \mbox{LTE-U} and \mbox{Wi-Fi} when all nodes can hear each other is modeled and studied in~\cite{WL},\cite{info}. But, in a general scenario, nodes are deployed spatially and all the nodes are not always inside EDT of each other. The placement of both \mbox{Wi-Fi} and \mbox{LTE-U} nodes can be quite arbitrary due to many practical constraints in the environment. Even multiple parallel LTE-U and Wi-Fi transmissions are possible on a given channel based on the deployment scenario. Dense deployment of \mbox{Wi-Fi} and \mbox{LTE-U} networks can have any kind of placement in indoor/outdoor scenarios. Hence, the throughput estimation for a generalized scenario is an essential requirement for the study of fair coexistence of \mbox{LTE-U} and Wi-Fi networks, and it can also help to do efficient placement of LTE-U and Wi-Fi nodes. In this paper, we propose an analytical model to estimate throughput of \mbox{LTE-U} and \mbox{Wi-Fi} networks in a generalized scenario. 

 \section{System model}  \label{SM}
We consider a scenario where multiple LTE-U eNBs and Wi-Fi APs are deployed in a spatially distributed scenario and operating on the same unlicensed channel as shown in Fig.~\ref{ExampleNet}a. In the figure, there are three \mbox{LTE-U} nodes (L1, L2, and L3) and three \mbox{Wi-Fi} nodes (W1, W2, and W3). The \mbox{LTE-U} eNBs and Wi-Fi APs can listen only a subset of nodes in the networks. The \mbox{LTE-U} nodes are following ON-OFF cycle for fair coexistence with \mbox{Wi-Fi}. The sum of LTE-U ON time and OFF time durations is the duty cycle period denoted by $T_{frame}$. \mbox{Wi-Fi} nodes are following the distributed CSMA/CA protocol in which they do a Clear Channel Assessment (CCA) before channel access. 
\subsection{Assumptions} \noindent
\textbf{1) Traffic:} We consider only the downlink traffic for both LTE-U and Wi-Fi networks. In addition, we assume a full-buffer case where each node (eNB/AP) always has data for transmission similar to~\cite{boe}.\\
\textbf{2) LTE-U Coordination:} \mbox{LTE-U} nodes can easily coordinate among themselves using X2-interface to prevent channel wastage due to collisions. Therefore an \mbox{LTE-U} can reschedule its transmission when another \mbox{LTE-U} is actively transmitting in its range. Hence, for better coordination, we assume that the duration of $T_{frame}$ is same for all the LTE-U nodes. Since each LTE-U node transmits for a duration equal to its LTE-U ON time in $T_{frame}$, we assume that the LTE-U nodes start their transmissions at the beginning of each $T_{frame}$. Therefore,  \mbox{LTE-U} nodes complete their transmissions at the beginning of $T_{frame}$ and provide an opportunity to \mbox{Wi-Fi} nodes during remaining duration of $T_{frame}$.
\subsection{LTE-U--Wi-Fi network graph}
The CSMA/CA protocol of Wi-Fi follows the LBT mechanism in which nodes perform CCA on the shared channel so that no two Wi-Fi nodes inside carrier sense range transmit simultaneously. A \mbox{Wi-Fi} node considers the channel as busy if it detects any Wi-Fi signal exceeding the Carrier Sense Threshold (CST), or if it detects any signal other than Wi-Fi exceeding the Energy Detection Threshold (EDT). In the network graph of Fig.~\ref{ExampleNet}b, an edge between two \mbox{Wi-Fi} nodes indicates that both the nodes are inside CST range of each other and an edge between Wi-Fi and LTE-U nodes or LTE-U and LTE-U nodes indicates that these nodes are inside EDT range of each other. Thus, whenever there is an edge between two nodes, only one of the nodes can successfully transmit on the channel whereas no edge between the nodes indicate that both the nodes can transmit simultaneously.

\subsection{LTE-U network graph}
Each LTE-U node performs channel load assessment during its OFF duration to tune its duty cycle for the next $T_{frame}$. Before beginning its transmission, each LTE-U node checks for the status of other LTE-U nodes in its range (through X2-interface). If no LTE-U node is actively transmitting, it starts its transmission. We draw a separate LTE-U network graph from Wi-Fi--LTE-U network graph as shown in Fig.~\ref{ExampleNet}c to derive the possible states of the LTE-U nodes. We note that though Wi-Fi information is missing in the LTE-U network graph, LTE-U node's ON duration is derived from the activity of all the nodes (Wi-Fi and LTE-U) in its vicinity. The LTE-U network graph is then used to draw the Wi-Fi contention graph which helps to estimate throughput of Wi-Fi nodes operating on the shared channel. 
     \begin{figure}[h]
     	\begin{center}
     		\includegraphics[width=7.5cm,height=2.6cm]{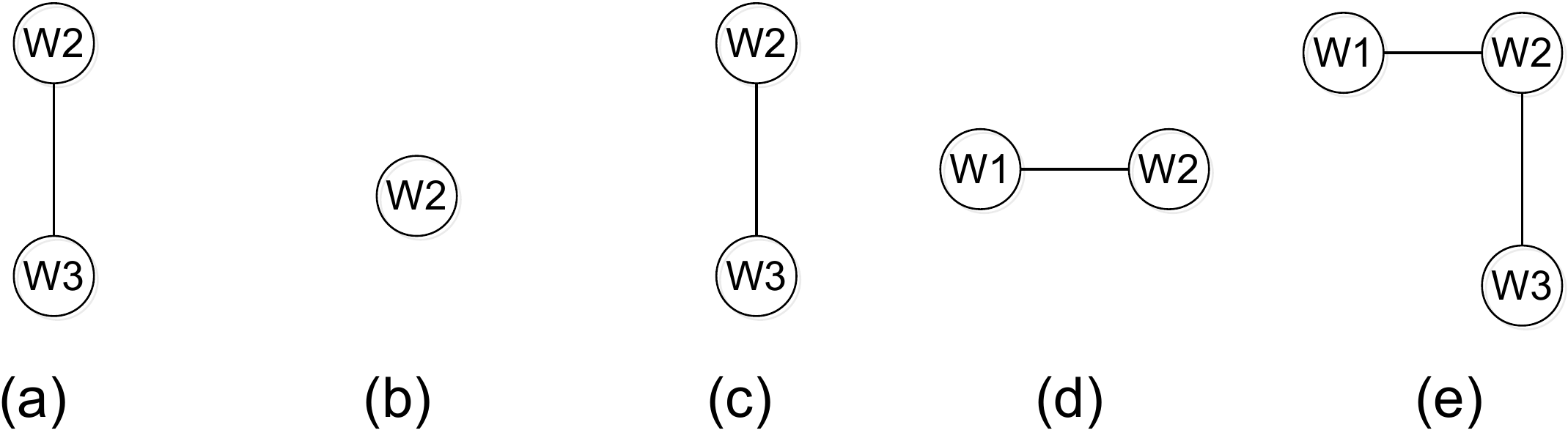}
     		\newline
     		\caption{Wi-Fi contention graphs corresponding to LTE-U states: (a) L1 and L2 are transmitting, (b) L1 and L3  are transmitting, (c) Only L1/L2 is transmitting, (d) only L3 is transmitting, (e) no LTE-U nodes are transmitting.  }
     		\label{S2}
     	\end{center}
     \end{figure}
\subsection{Wi-Fi contention graph}
LTE-U network graph serves as a basis for formulating Wi-Fi contention graphs. Wi-Fi nodes are opportunistic unlike LTE-U nodes which are dominant in channel access. Wi-Fi nodes do CCA before transmitting; whenever there is any ongoing transmission of LTE-U node, Wi-Fi nodes cannot participate in contention for the duration of LTE-U ON. Therefore, we derive the Wi-Fi contention graphs as shown in Fig.~\ref{S2} with respect to \mbox{LTE-U} states as governed by LTE-U network graph shown in Fig.~\ref{ExampleNet}b. In Fig.~\ref{S2}a, we can see the Wi-Fi contention graph when nodes $L1$ and $L2$ are transmitting. Here, node $W1$ is suppressed by transmissions from $L1$ and $L2$ and it cannot participate in contention.  Similarly, based on states of \mbox{LTE-U} nodes in the network, other possible Wi-Fi contention graphs can be derived as shown in Figs.~\ref{S2}b to~\ref{S2}e. In the next section, we will derive the LTE-U states from LTE-U network graph and then build the corresponding \mbox{Wi-Fi} contention graphs.

\section{Problem Formulation: Throughput computation from LTE-U state diagram}\label{Motivation}
In this section, we illustrate the behavior of \mbox{Wi-Fi} and \mbox{LTE-U} networks by considering the example network scenario given in the system model. In the later sections, we use these observations to model the achievable throughputs of \mbox{LTE-U} and \mbox{Wi-Fi} networks.
\subsection{Behavior of LTE-U and Wi-Fi networks in coexistence scenario}
We study the LTE-U--Wi-Fi coexistence behavior using the example network scenario shown in Fig.~\ref{ExampleNet}a. The state of each \mbox{LTE-U} node in the network could be one among $\{0,1,2\}$ which mean \textit{yet to transmit}, \textit{transmitting}, and \textit{completed transmission}, respectively.

\par From the design point of \mbox{LTE-U} protocol, an \mbox{LTE-U} node decides its duty cycle based on the load it senses on the channel. In the saturation case, load can be roughly mapped to the number of other nodes in its contention region. Hence, L2 and L3 nodes in Fig.~\ref{ExampleNet} are surrounded by same number of nodes and they tune their duty cycle duration for almost same time. Similarly, L1 decides its duty cycle based on the number of other nodes in its vicinity. Here we note that the duty cycle is solely derived from LTE-U protocol consideration, but to keep it simple, we are going with the number of nodes in its vicinity.
\par
\begin{figure}[h]
	\begin{subfigure}{\linewidth}
		\includegraphics[width=\linewidth]{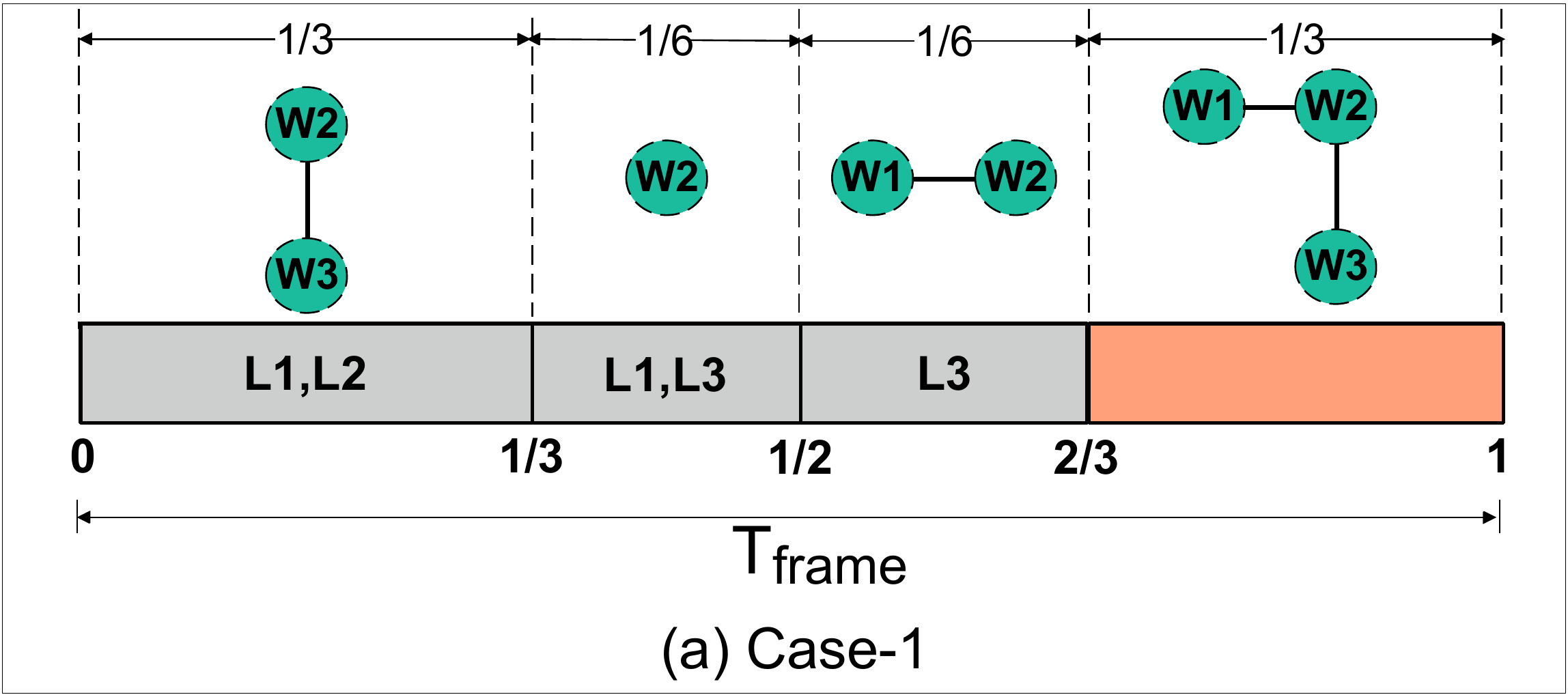}

		\label{C1}
	\end{subfigure}
	\newline
	\begin{subfigure}{\linewidth}		\includegraphics[width=\linewidth]{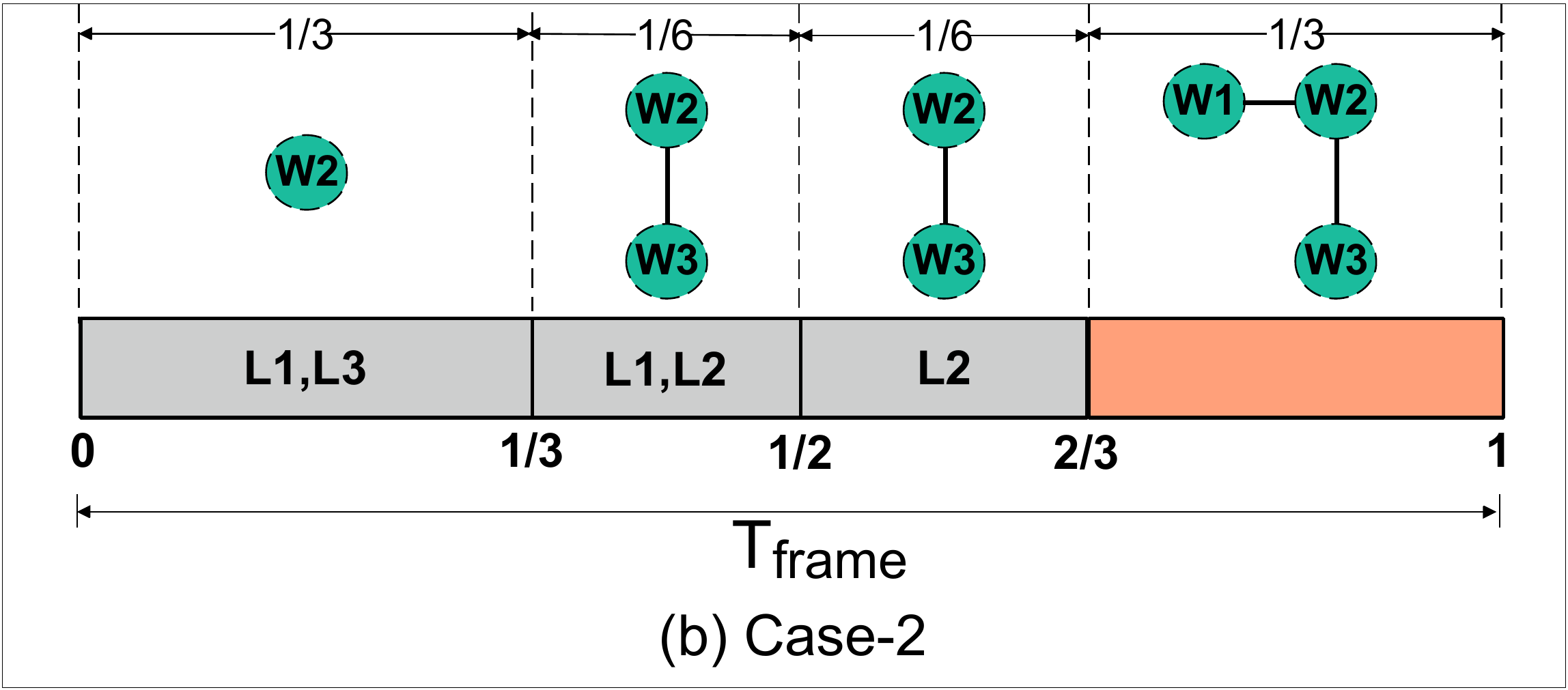}
	
		\label{C2}
	\end{subfigure}
	\caption{LTE-U transmission cases and change in corresponding Wi-Fi contention graphs, over $T_{frame}$ for (a) Case-1 (b) Case-2.}
	\label{3}
\end{figure}

 In the system model, we considered that LTE-U nodes transmit at the beginning of $T_{frame}$ and allow Wi-Fi nodes to transmit later. So, for the network given in Fig.~\ref{ExampleNet}a, we can draw the time sequence diagram of the network in two ways depending on which node transmits first; Case-1: L2 transmits first, Case-2: L3 transmits first, as shown in Fig.~\ref{3}. In both the cases, L1 along with L2/L3 can transmit simultaneously, because it is outside the range of L2 and~L3, as observed from the LTE-U network graph in Fig~\ref{ExampleNet}c.
In Fig.~\ref{3}, $T_{frame}$ is divided into several time segments based on the transmissions from \mbox{LTE-U} nodes in the network. The \mbox{LTE-U} nodes L1, L2, and L3 are specified inside the box when they are transmitting in the corresponding time segment, and accordingly the \mbox{Wi-Fi} contention graph is shown above the corresponding time segment. In each time segment, the \mbox{Wi-Fi} contention graph consists of only those Wi-Fi nodes which can successfully transmit in that time segment while those Wi-Fi nodes which cannot transmit because of LTE-U transmissions are excluded from the Wi-Fi contention graph. From Fig.~\ref{3}, we observe two major factors while calculating throughput in the coexistence scenario of \mbox{LTE-U} and \mbox{Wi-Fi} networks:\\
\textbf{1) Dynamic changes in \mbox{Wi-Fi} contention graph:}
Changes in \mbox{LTE-U} node's transmission affect the \mbox{Wi-Fi} contention graph over $T_{frame}$.  
\textbf{2) Dependency of \mbox{Wi-Fi} contention graph on \mbox{LTE-U} state transition:} We found that the \mbox{Wi-Fi} contention graph depends on state transition of \mbox{LTE-U} nodes.
\par The above-mentioned factors are shown for the example network shown in Fig.~\ref{ExampleNet}a. In each case, we can see a frequent change in \mbox{Wi-Fi} contention graph, depending on state of \mbox{LTE-U} nodes in the network. The dependency of \mbox{Wi-Fi} contention graph on the state transition of \mbox{LTE-U} nodes is clear with two cases as shown in Fig.~\ref{3}. We can observe that while in Case-1, W1 is getting an opportunity to transmit for 1/2 fraction of $T_{frame}$, in Case-2, W1 is getting an opportunity to transmit for 1/3 fraction of $T_{frame}$ which results in different throughputs of \mbox{Wi-Fi}. We formulate a mathematical model in the next section for the LTE-U state network diagram which helps in organizing the computation procedure.
\begin{figure}[h]
	\begin{center}
		\includegraphics[width=\linewidth]{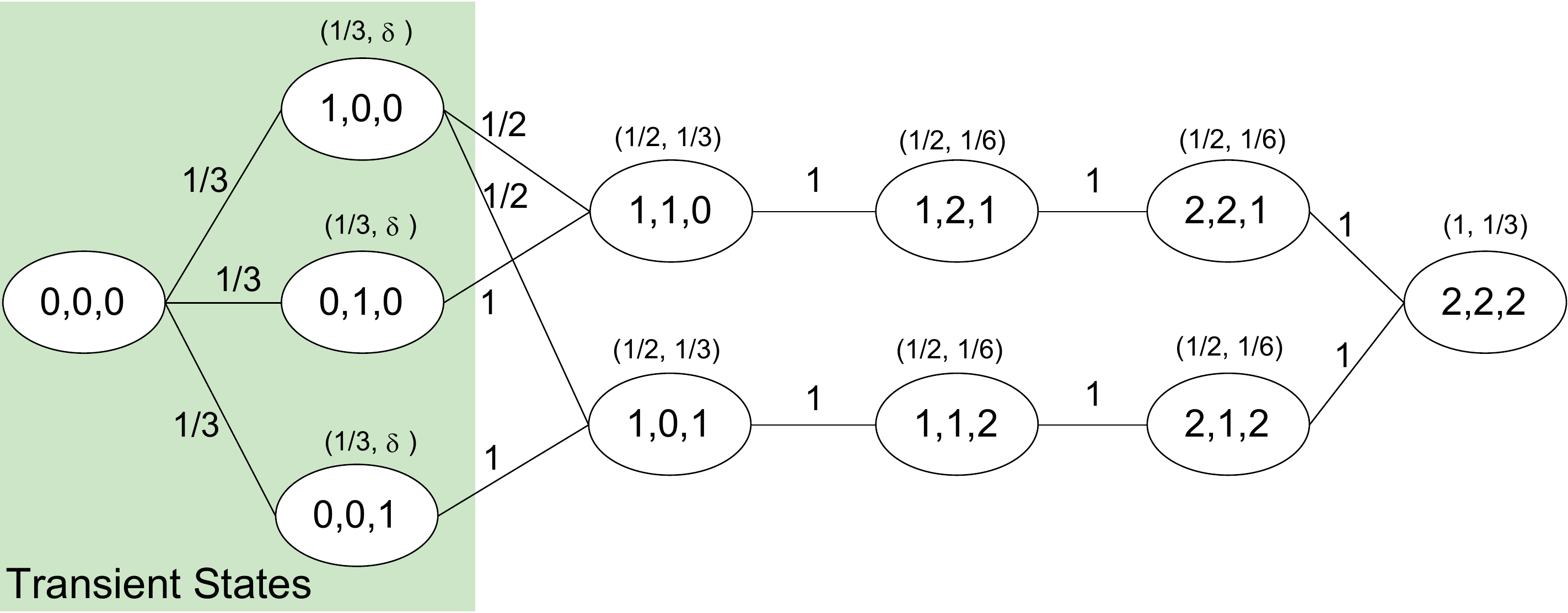}
		\caption{LTE-U network state transition diagram of L1, L2, and L3 for the example network shown in Fig.~\ref{ExampleNet}a.}
		\label{Exstate}
	\end{center}
\end{figure}
\subsection{Throughput computation from LTE-U network state transition diagram}
Fig.~\ref{Exstate} shows the LTE-U network state transition diagram, in which each system state is formed by considering the current state of each of the three \mbox{LTE-U} nodes which is one among $\{0,1,2\}$ as discussed earlier. LTE-U network states are valid if and only if no two \mbox{LTE-U} nodes within EDT range are in State 1 simultaneously. From Fig.~\ref{Exstate}, we can observe how the \mbox{LTE-U} network state gets updated with time as the transmission time of \mbox{LTE-U} node gets exhausted. There are different possible paths from a state because of the independent behavior of \mbox{LTE-U} nodes. At a given point of time, there is an equal probability for any \mbox{LTE-U} node to start its transmission. Each path is associated with certain probability, and all the paths have equal probability to occur because of the independent nature of LTE-U nodes. The probability sum gained in each path (previous state probability $\times$ probability of path) leading to a system state gives the probability of that particular state. Each state occurs for a certain amount of time, and then moves on to the next state when any \mbox{LTE-U} node changes its state. Time spent in each state is minimum of the leftover transmission times of \mbox{LTE-U} nodes. In Fig.~\ref{Exstate}, the probability of occurrence of a state and the time spent in that state are marked respectively on the top of each state. Further, we can see that some states are marked as transient states. They are called transient states because they exist for negligible amount of time ($\delta$ denotes negligible amount of time in Fig.~\ref{Exstate}) and it is possible for some other \mbox{LTE-U} node to change its state. For instance, consider the first transient state $(1,0,0)$ in the $2^{nd}$ level of the state transition diagram. There is a scope for $L2$ or $L3$ to start its transmission, so they quickly slide to upcoming states $(1,1,0)$ or $(1,0,1)$ where they spend some time and then move on to the next state. Though transient states are insignificant with respect to time, they help in computing the probability of upcoming states. After some time, all \mbox{LTE-U} nodes complete their transmissions and the system will be in state $(2,2,2)$. Here onwards \mbox{LTE-U} nodes wait for $T_{frame}$ to exhaust to refresh their states to $(0,0,0)$. After having this state diagram, if we examine the behavior of \mbox{Wi-Fi}, we will observe that \mbox{Wi-Fi} (being CSMA based technology) looks for channel status before transmitting and in a given state of \mbox{LTE-U} network some \mbox{Wi-Fi} nodes are completely dominated because of the transmissions of \mbox{LTE-U} nodes in their range. We would get a \mbox{Wi-Fi} contention graph corresponding to each system state as shown in Fig.~\ref{3}. Throughput is computed in each case for getting overall system throughput. With this motivation, we try to generalize our state model construction, but instead of considering discrete state space we consider continuous state space. In the next section, we describe a generalized way of constructing continuous state space \mbox{LTE-U} network and computing the corresponding state probabilities. Further, this information is used to model the throughput of \mbox{Wi-Fi}.
\section{Analytical model for Throughput Estimation}\label{model}
Based on the system model described in the previous section, we now present a generalized way to model the throughput of spatially distributed \mbox{Wi-Fi}--\mbox{LTE-U} networks. Since the network behavior is periodic (with a time period of $T_{frame}$), a couple of events may change in every $T_{frame}$, leading to a set of cases as described in the previous section. But, after a considerable amount of time, we can safely assume that those events are equiprobable to occur, which exhibit equal effect on the network.

Notation used in the analytical model are given in Table~\ref{notations}.

\begin{table}[htb!]
	\caption{Notation used for LTE-U and Wi-Fi modeling}
	\centering
	\begin{tabular}{|p{1.5cm}| p{5.8cm}|}
		\hline\bfseries
		Notation&\bfseries \hspace{2.5cm} Definition \\
		\hline \hline
		$N_{L}$& Number of LTE-U nodes in the network \\
		\hline
		$N_{W}$& Number of Wi-Fi nodes in the network\\
				\hline
		$T_{frame}$& LTE-U ON + LTE-U OFF time (same for all LTE-U nodes in the network)  \\
		\hline
		$S_{l}$& $l^{th}$ state of LTE-U network in the transition model   \\
		\hline
		$S_{l}^{i}$ & The state that $i^{th}$ LTE-U node is supposed to be in system state $S_{l}$ \\
		\hline
		$\psi_{S_{l}}(t)$& Probability of state $S_{l}$ at time $t$ \\
		\hline
		$P_{l}$ & Set of paths through which we can reach state $S_{l}$ \\
		\hline
		$\Upsilon_{S_{l}}$& Time at which path set $P_{l}$ is non-empty in reaching state $S_{l}$ \\
		\hline
		$\chi_{S_{l}}^{i}$& State of $i^{th}$ LTE-U node when system is in state $S_{l}$, $\chi_{S_{l}}^{i}$ $\in \{0,1,2\}$ \\
		\hline
		$\tau_{S_{l}}^{i}(t)$& Left over transmission time of $i^{th}$ LTE-U node in state $S_{l}$ at time t\\
		\hline
		$f()$& Function which decides $T_{ON}$ of LTE-U node based on LTE-U protocol standard\\
		\hline
		$D^{i}$& Duty cycle of $i^{th}$ LTE-U node  \\
		\hline
		$\psi_{i}$& Probability of $i^{th}$ path \\
		\hline
		$O_{S_{l}}$& Number of outlets from sate $S_{l}$\\
		\hline
		$g(W_{i},S_{l})$& State of $i^{th}$ Wi-Fi node, when LTE-U network is in state $S_{l}$, $g\in\{0,1\}$ \\
		\hline
		$W_{a}(S_{l})$& Set of Wi-Fi nodes which can actively transmit in LTE-U system state $S_{l}$ \\
		\hline
		$L_{set}$ & Set of all feasible LTE-U network states with a non-zero probability at some point of time between 0 to $T_{frame}$\\
		\hline
		EDT(i) & Represents the set of LTE-U nodes in the EDT region of Wi-Fi/LTE-U node $i$\\
		\hline
		CST(i) & Represents the set of Wi-Fi nodes in the EDT region of Wi-Fi $i$\\
		\hline
		$\psi_{i}^{k}$ & Denotes the probability of $i^{th}$ LTE-U node to be in state $k$ (where $k$ $\in$ 0,1,2)\\ 
		\hline
		BOE() & Gives the throughput estimation of Wi-Fi nodes using BOE model\\
		\hline
		$A$  & Air time fraction, it is a ratio of channel access time (successful + collision) to the total air time.\\
		\hline
		$A_{W}^{i}$ & Air time fraction of $i^{th}$ Wi-Fi node\\
		\hline
    	$A_{L}^{i}$ & Air time fraction of $i^{th}$ LTE-U node\\
    	\hline
		$Q_{i}(S_{l})$& $i^{th}$ Independent state of Wi-Fi network formed when LTE-U network state is $S_{l}$ \\
		\hline
		$IS(S_{l})$& Set of Independent States of Wi-Fi network in LTE-U system state $S_{l}$ \\
		\hline
		$MIS(S_{l})$& Set of Maximum Independent States of \mbox{Wi-Fi} network in LTE-U system state $S_{l}$, $|\ MIS(S_{l})\ |=N_{MIS}(S_{l})$ \\
		\hline
		$W_{nor}(S_{l})$& Normalization throughput distribution of Wi-Fi network in state $S_{l}$. $W_{nor}$ is averaged normalization throughput distribution of Wi-Fi network over all LTE-U network states\\
		\hline
	\end{tabular}
	\label{notations}
\end{table}

\subsection{LTE-U Network State Transition Model}
\begin{figure}
	\begin{center}
		\includegraphics[width=8.7cm,height=4.0cm]{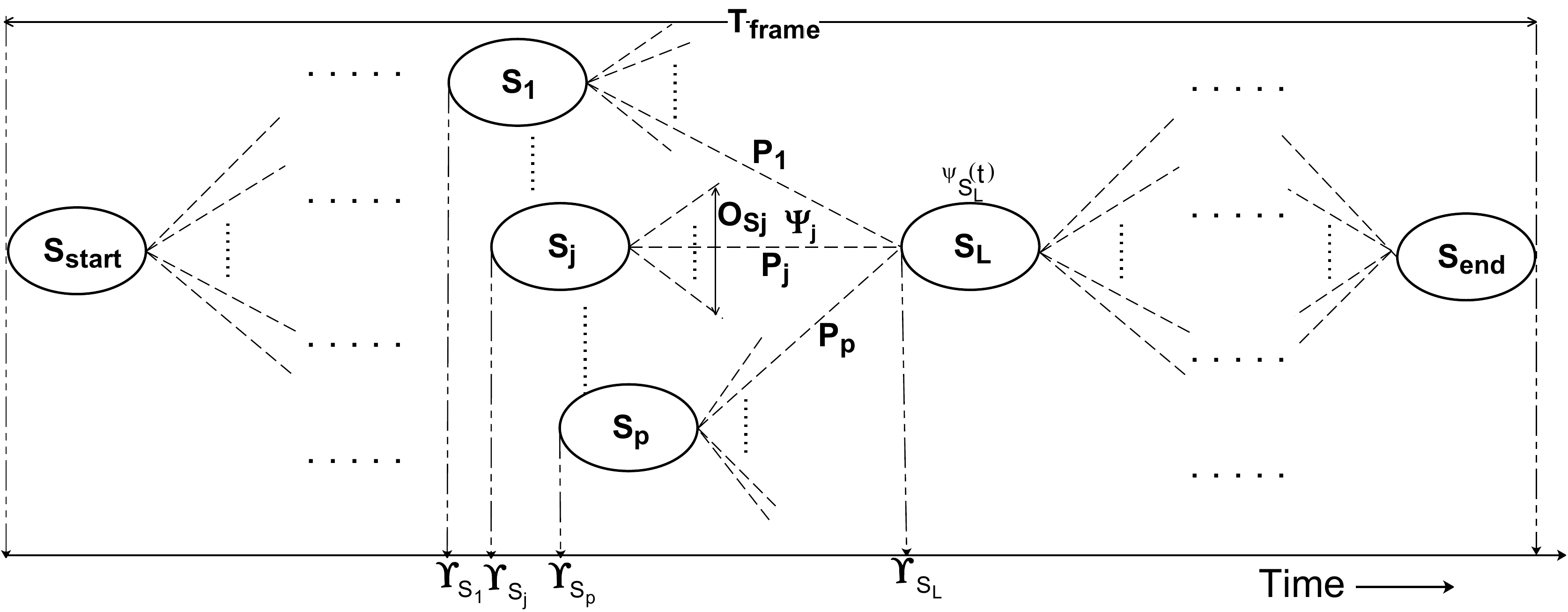}
		\caption{LTE-U network state transition diagram.}
		\label{Gstate}
	\end{center}
\end{figure}

Let $N_{L}$  be the number of \mbox{LTE-U} eNBs in the network and $S_{l}$ be the set which describes the~\mbox{LTE-U} network state. $S_{l}=\{S_{l}^{1},\cdots,S_{l}^{i},\cdots,S_{l}^{N_{L}}\}, i \in (1,N_{L})$. The set of all feasible LTE-U network states is given by $L_{set}=\{S_{1},\cdots,S_{l},\cdots,S_{M}\}$, where M is the number of feasible LTE-U network states. Since each LTE-U network state $S_{l}$ occurs at different point of time during LTE-U network state transition, the probability associated with the LTE-U network state occurrence changes with time as $\psi_{S_{l}}(t)$.
$S_{l}^{i}$ is an ordered pair used to describe the LTE-U state of $i^{th}$ LTE-U node when LTE-U network is in state $S_{l}$. $S_{l}^{i}=(\chi_{S_{l}}^{i},\tau_{S_{l}}^{i}(t))$, where $\chi_{S_{l}}^{i} \in \{0,1,2\}$ and $\{0,1,2\}$ represent \textit{yet to transmit, transmitting, and completed transmission,} respectively. $\tau_{S_{l}}^{i}(t)$ represents the left over transmission time of an \mbox{LTE-U} node $L^{i}$ when LTE-U network is in state $S_{l}$. $\tau_{S_{l}}^{i}(t)$ decreases at a rate of $\frac{d(\tau_{S_{l}}^{i}(t))}{dt}=-1$ if $\chi_{S_{l}}^{i}=1$ and $\frac{d(\tau_{S_{l}}^{i}(t))}{dt}=0$ for nodes with $\chi_{S_{l}}^{i}=0\ or\ 2$. The LTE-U network state, except in the stable state, continuously changes as \mbox{LTE-U} transmission count drops linearly with time. Fig.~\ref{Gstate} represents a general LTE-U network transition diagram. Time is on X-axis and states are placed in a two-dimensional space corresponding to their occurrence times as marked by $\Upsilon_{S_{j}}$ for $j^{th}$ LTE-U network state. All LTE-U network states are connected with dotted lines for representing state transitions. In each $T_{frame}$, the LTE-U network starts at $S_{start}$, (\emph{i.e.,} all LTE-U nodes are waiting for their opportunity to transmit) and through series of transitions they reach to $S_{end}$ which represents that all LTE-U nodes have finished their transmissions. The left over air time is given for Wi-Fi till another $T_{frame}$ starts. This LTE-U network state transition diagram forms a basis for analysing the network. During the transition of LTE-U network, it goes through a series of transient states and finally reaches to the stable state. The terms transient and stable states are defined below. \\
\textbf{Definitions:} \\
\textbf{Stable State:} An LTE-U network state $S_{l}$ is considered to be stable when $\forall\ S_{l}^{i}\in S_{l}\ |\ \chi_{S_{l}}^{i}=2$, \emph{i.e.,} all \mbox{LTE-U} nodes have completed their transmissions ($\tau_{S_{l}}^{i}=0$) and LTE-U network state transition has come to an end.\\
\textbf{State Connectivity:} As shown in Fig.~\ref{Gstate}, transitions in our state transition model are unidirectional. It starts at time $t=0$ and always propagates forward towards a stable state through a series of states. A transition is possible if and only if the states are connected. Two LTE-U network states $S_{1}$ and $S_{2}$ are said to be connected when there is at most one \mbox{LTE-U} node for which $\chi_{S_{1}}^{i}\neq \chi_{S_{2}}^{i}$ and after $\Delta t$ time state $S_{1}$ converges to state $S_{2}$. $S_{1}$ and $S_{2}$ are the left and right states, respectively.\\
\textbf{Transient states:} Here we define transient states in much greater detail, which are observed in the state transition diagram when after $\Delta t$ amount of time, there is a possibility for one or more \mbox{LTE-U} nodes to change their $\chi_{S_{l}}^{i}$. In such cases, the network enters into a transient state, and according to the definition of state connectivity no more than one \mbox{LTE-U} state can change in one state transition. Therefore, the network makes series of state transitions to come out of transient states and resumes its normal countdown. Transient states occur when the network is about to make a transition which involves the change in $\chi_{S_{l}}^{i}$ from 0 $\rightarrow$ 1 or 1 $\rightarrow$ 2. A transition from 0 $\rightarrow$ 1 will occur when there exists an \mbox{LTE-U} node which has not yet transmitted and has no \mbox{LTE-U} node which is transmitting in its region. This can be expressed by a conditional statement as follows: $S_{l}\ |\ \exists L^{i},\chi_{S_{l}}^{i}=0 \land \forall L^{j} \in EDT(L^{i}), \chi_{S_{l}}^{j}=\{0,2\}$. A transition from 1 $\rightarrow$ 2 can be identified as follows, for state $S_{l}\ |\ \tau_{S_{l}}^{i}(t)=\Delta t$. $\Delta t$ is very small and after $\Delta t$ amount of time, the state is going to be transient awaiting for state transition from 1 $\rightarrow$ 2. The system quickly slides to next state from the transient state. Though transient states are not significant with respect to time and throughput computation, they decide the probability of upcoming state and holds the definition of state connectivity. 

\subsubsection{LTE-U Throughput Modeling}
\mbox The LTE-U nodes, based on the activities on the channel, decides the LTE-U ON period ($T_{ON}$) in each $T_{frame}$. Let $f()$ be the function (derived from LTE-U protocol) that assigns the LTE-U node with some specific $T_{ON}$ during the starting of $T_{frame}$, $\tau_{S_{start}}^{i}(0)=f(L^{i})$. At any time $t$, $T_{ON}$ is given as $\tau_{S_{l}}^{i}(t)$ which decreases at a rate of $\frac{d(\tau_{S_{l}}^{i}(t))}{dt}=-1$ if \mbox{LTE-U} node is transmitting or 0 if it has completed transmission or yet to transmit. $D^{i}$, duty cycle of \mbox{LTE-U} node $L^{i}$ is given as
\begin{equation}
D^{i}=\frac{f(L^{i})}{T_{frame}}
\end{equation}
Having known the duty cycle of LTE-U node, the LTE-U throughput can be computed by multiplying duty cycle with basic data rate of LTE-U. Let basic data rate of LTE-U node be $\sigma_{l}$, then the throughput of $i^{th}$ LTE-U node is given as
\begin{equation}
L_{Thr}^{i}=D ^{i} \times \sigma_{l}
\end{equation}
\subsubsection{State Probability}
Let $P_{l}$ denote the set of paths through which we can reach an \mbox{LTE-U} network state $S_{l}$, and the time at which the set $P_{l}$ is non empty be $\Upsilon_{S_{l}}$. Let $|P_{l}|$ be $p$, $P_{l}$ can be represented as  $P_{l}=\{P^{j}\}, j\in(1,p)$. Let $S_{j}$ be the corresponding previous LTE-U network state of $j^{th}$ path. $\psi_{S_{j}}(t)$ denotes the probability of occurrence of LTE-U network state  $S_{j}$, and $\psi_{j}$ denotes the probability of the path $j$ as shown in Fig.~\ref{Gstate}. Since \mbox{LTE-U} state transitions are independent as it is not following any backoff mechanism, each state transition is equiprobable to occur. $\psi_{j}$ can be represented in terms of the number of states $S_{j}$ can lead to after $\Delta t$. Let $O_{S_{j}}$ represent the number of states $S_{j}$ can transit to as shown in Fig.~\ref{Gstate}. Therefore, $\psi_{j}=\frac{1}{O_{S_{j}}}$. Thus, $\psi_{S_{l}}(\Upsilon_{S_{l}})$ can be computed as
\begin{equation}
	\psi_{S_{l}}(\Upsilon_{S_{l}})=\sum_{j=1}^{p} \psi_{j}\  \times \psi_{s_{j}}(\Upsilon_{S_{l}}-\Delta t)
\end{equation}
After calculating the probability of state $S_{l}$ at time $t=\Upsilon_{S_{l}}$, we can calculate $\psi_{S_{l}}$ at any time $t$ after $\Upsilon_{S_{l}}$ from Eqn.~(\ref{n2}).
\begin{equation}\label{n2}
\psi_{S_{l}}(t)=\psi_{S_{l}}(\Upsilon_{S_{l}})+\int_{\Upsilon_{S_{l}}}^{t}d(\psi_{S_{l}}(t))
\end{equation}
\newline
The remainder of this section derives the probability of a particular LTE-U network state at time $t$ in terms of probability of LTE-U network states at $t-\Delta t$. We derive the set of paths $P_{l}$ and dissolve previous state probability $\psi_{S_{j}}(t-\Delta  t)$ into newer state. According to state connectivity, two LTE-U network states are connected by at least one path if and only if after  $\Delta t$ amount of time, LTE-U network state $S_{j}$ converges to LTE-U network state $S_{l}$. Hence, we see that the state $S_{j}$ contributes in building the probability of state $S_{l}$ at time $t$. Events that are responsible for building the probability of an LTE-U network state are explained next. 
\par
\begin{figure}[h]
	\begin{center} 
		\includegraphics[width=\linewidth]{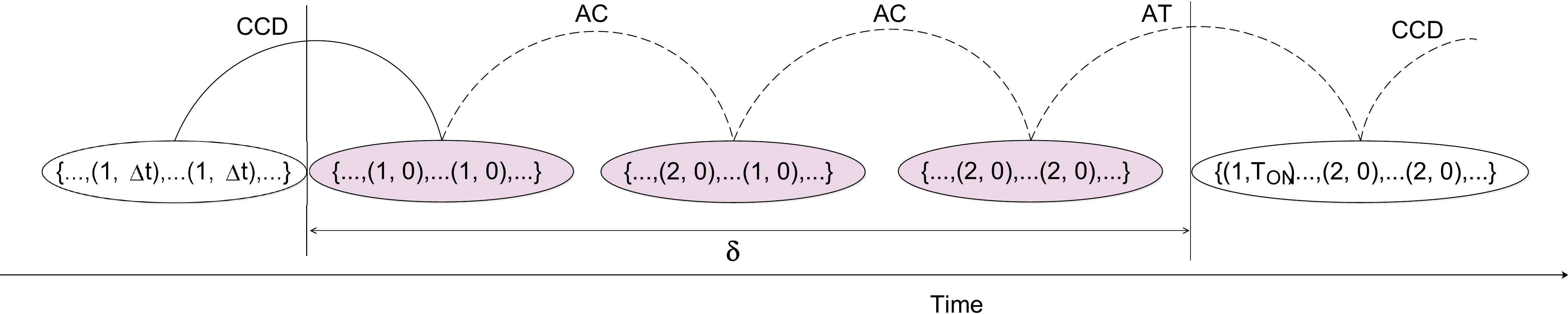}
		\caption{Event occurrence order in LTE-U network state  transition diagram.}
		\label{events}
	\end{center}
\end{figure}

 LTE-U network states having at least one \mbox{LTE-U} node which has exhausted its transmission count waits in a transient state to move from 1 $\rightarrow$ 2. Such kind of state changes only occur from transient states. This particular event is denoted by $AC$ (\emph{i.e.,} About to Complete). Another event that is responsible for probability contribution is transition from 0 $\rightarrow$ 1. Even this state change occurs from a transient state and is denoted by $AT$ (\emph{i.e.,} About to Transmit). Final event which contributes in building up the probability of an LTE-U network state is continuous countdown of transmission time of actively transmitting \mbox{LTE-U} nodes, which is denoted as $CCD$ (\emph{i.e.,} Continuous Countdown). The LTE-U network spends most of the time in $CCD$ and finally reaches the stable state where all transmissions have stopped. Even the stable state is counted under $CCD$ but there are no \mbox{LTE-U} nodes which are left for countdown. It is not necessary for all the above stated events to contribute to LTE-U network state probability. Some of the mentioned events are not applicable for some LTE-U network states. For instance consider an LTE-U network state where \mbox{LTE-U} has just started transmission. In this state, $CCD$ does not contribute to the probability because \mbox{LTE-U} transmission time cannot be greater than the time it has when it just begins transmission ($D^{i}$ $\times$ $T_{frame}$). Therefore, contribution of $CCD$ for such \mbox{LTE-U} network states is zero. The occurrence of the above described events in the sequence is shown in Fig.~\ref{events}. The events $AC$ and $AT$ which occur from transient states are separated by borders from normal LTE-U network states. The dotted line in Fig.~\ref{events} represents the possibility of multiple transitions. In the figure, only a few states are shown for explanation. It is not always necessary that $AT$ event occurs after $AC$. The system may directly go to the $CCD$ event. Such a phenomena depends on whether there still exists an \mbox{LTE-U} node which has not yet transmitted. However, it is always guaranteed that a transient state occurs before $AT$ event. LTE-U network spends all its time in CCD events, this follows by the fact that AT and AC events deal with the transient states which consume a negligible amount of air time.
\newline
With the knowledge of the events that occur during state transition of LTE-U network, $\psi_{S_{L}}(t)$ can be  derived from previous state probabilities as follows: 
\begin{equation}
\psi_{S_{l}}(t)=AC+AT+CCD
\end{equation}
\textbf{AC Event:}
Let $C$ denote the set of \mbox{LTE-U} nodes which have completed transmissions ($\chi_{S_{l}}^{i}=2$) according to \mbox{LTE-U} network state $S_{l}$  and $C^{i}$ be the $i^{th}$ element of the set $C$. $M$ is a set of nodes which are actively transmitting according to state $S_{l}$. Then the probability contribution of AC event is derived as follows:
\begin{equation}\label{e6}
	AC=\sum_{i=1}^{|C|} \psi_{(\chi^{1}_{C^{i}}.\tau^{\delta}_{C^{i}}.\tau_{M}.S_{l})}(t-\delta) \times \frac{1}{N_{i}}
\end{equation}
where $\delta<<\Delta t$, because transient states are virtual and they do not consume any air time. The state $S_{l}$ is been acted upon by series of operators to transform it to state $S_{j}$ which we are interested in at time $t-\delta$. State $S_{j}$ is responsible for $AC$ event. $\chi^{1}_{C^{i}}$ is an operator which sets $\chi_{S_{j}}^{C^{i}}$ of \mbox{LTE-U} node $C^{i}$ in state $S_{j}$ to 1 and $\tau^{\delta}_{C^{i}}$ is an operator which sets $\tau_{S_{j}}^{C^{i}}(t-\delta)$ of \mbox{LTE-U} node $C^{i}$ to (0, $\delta$), where $\tau_{S_{j}}^{C^{i}}(t-\delta)$ should lie in the range of $(0,\delta)$ as $max(\tau_{S_{j}}^{C^{i}}(t-\delta))=(|\frac{d(\tau_{S_{j}}^{C^{i}}(t))}{dt}|_{|t-\delta}) \times  \delta=\delta$. $\tau_{M}$ is an operator which operates on transmission time of \mbox{LTE-U} nodes which are actively transmitting according to state $S_{l}$ (which is given by set $M$), $\tau_{S_{j}}^{M^{i}}(t-\delta)=\tau_{S_{l}}^{M^{i}}(t)+\Delta^{i},i\in (1,|M|)$ and $\Delta^{i}\in (0,\delta)$, it follows from $\Delta^{i}\leq(|\frac{d(\tau_{S_{j}}^{M^{i}}(t))}{dt}|_{|t-\delta}) \times \delta$. The fraction $\frac{1}{N_{i}}$ appears in Eqn.~(\ref{e6}) because multiple nodes can end transitions at once which results in various possible paths with each path equiprobable to occur. $N_{i}$ can be derived as follows:
\begin{equation}
N_{i}=|\ \{S_{j}^{i}\ |\ \chi_{S_{j}}^{i}=1\ \land\ \tau_{S_{j}}^{i}(t-\delta)\in (0,\delta)\}\ |
\end{equation}
\textbf{AT Event:}
Let $M$ denote the set of \mbox{LTE-U} nodes which are actively transmitting in state $S_{l}$ and $M^{k}$ denote $k^{th}$ element of set $M$. Probability contribution of $AT$ event is derived as follows:
\begin{equation}\label{e7}
	AT=\sum_{k=1}^{|M|}\psi_{(\chi^{0}_{M^{k}}.\tau_{M}.S_{l})}(t-\delta) \times \frac{1}{N_{k}}
\end{equation}
Where $\delta<<\Delta t$ because of the characteristics of transient state. After operating state $S_{l}$ with the set of operators, the state transforms to $S_{j}$ which contributes to $AT$ event. $\chi^{0}_{M^{k}}$ is an operator that sets $\chi_{S_{j}}^{M^{k}}$ of \mbox{LTE-U} node $M^{k}$ in state $S_{j}$ to 0 and operator $\tau_{M}$ sets $\tau_{S_{j}}^{M^{i}}(t-\delta)=\tau_{S_{l}}^{M^{i}}(t)+\Delta^{i},i\in (1,|M|)$ and $\Delta^{i}\in (0,\delta)$, it follows from $\Delta^{i}\leq(|\frac{d(\tau_{S_{j}}^{M^{i}}(t))}{dt}|_{|t-\delta}) \times \delta$. The fraction $\frac{1}{N_{k}}$ appears in Eqn.~(\ref{e7}) due to various possible transition states $S_{j}$ (transient state) can be undergo one at time $t-\delta$. Note that all those state transitions are equiprobable. $N_{k}$ gives the number of \mbox{LTE-U} nodes that can possibly start transmission from transient state $S_{j}$ which can be derived as follows:
\begin{equation}
N_{k}=|\ \{\ S_{j}^{k}\ |\ \chi_{S_{j}}^{k}=0 \land \forall L^{m}\in EDT(L^{k}), \chi_{S_{j}}^{m}\in \{0,2\}\ \}\ |
\end{equation}
\textbf{CCD Event:}
A state $S_{j}$ can move to state $S_{l}$ as \mbox{LTE-U} transmission time drops off with time. State $S_{j}$ to reach $S_{l}$ at time $t$, it should satisfy the following condition: $\forall\ S_{l}^{i}\ |\ \chi_{S_{l}}^{i}=1, \tau_{S_{j}}^{i}(t-\Delta t)=\tau_{S_{l}}^{i}(t)+\Delta^{i}$, where $\Delta^{i}\in (0,\Delta t)$ and for nodes with $\chi_{S_{l}}^{i}=\{0,2\}$, it should follow that $\chi_{S_{j}}^{i}=\chi_{S_{l}}^{i}$. So state $S_{j}$ can be written in terms of state $S_{l}$ as $S_{j}$=$S_{l}+ds$. Probability gain incurred by event $CCD$ can be given as
\begin{equation}
	CCD=\psi_{(S_{l}+ds)}(t-\Delta t)
\end{equation}
For the above transition to happen after $\Delta t$ amount of time, $\Delta t \geq (|\frac{d(\tau_{S_{j}}^{i}(t))}{dt}|_{|t-\Delta t}) \times \Delta^{i}$.

\subsubsection{LTE-U node state prediction}
Now we can derive the corresponding LTE-U network states at a particular time $t$ and probability with which a particular LTE-U network state occurs. One interesting result that we can get with the current results is the prediction of state of \mbox{LTE-U} node in the network at a time $t$. At any time $t$, $\chi_{L}^{i}$ will be one among $\{0,1,2\}$. Therefore $\psi^{0}_{i}+\psi^{1}_{i}+\psi^{2}_{i}=1$, where $\psi^{k}_{i},k\in \{0,1,2\}$ denotes the probability with which \mbox{LTE-U} node $i$ be with $\chi_{L}^{i}=k$. Let  $L^{t,k}_{set}$ denote the set of LTE-U network states at time $t$ in which $\chi_{S_{l}}^{i}=k$, then $\psi_{i}^{k}$ is given as
\begin{equation}\label{probpred}
	\psi^{k}_{i}=\sum_{s=L_{set}^{t,k}} \psi_{s}(t)
\end{equation}
The above derived formula is validated later with simulation results in Section~\ref{PE}.
\subsection{\mbox{Wi-Fi} Throughput Modeling}
As mentioned earlier, \mbox{Wi-Fi} contention graphs are updated frequently as \mbox{LTE-U} network makes state transitions. \mbox{Wi-Fi} throughput needs to be modeled in each state of \mbox{LTE-U} network separately. In the literature, a lot of studies are made on the \mbox{Wi-Fi} network modeling. Here, we make use of Back-of-the-Envelop (BOE)~\cite{boe} throughput estimation technique to model the \mbox{Wi-Fi} throughput in LTE-U--Wi-Fi scenarios. BOE gives a rough estimation of \mbox{Wi-Fi} throughput which the authors demonstrated to be accurate enough for practical purposes when the number of nodes are below 50. \\
Let $N_{W}$ be the number of \mbox{Wi-Fi} nodes in the network. Let $g(W^{k},S_{l})$ be the function which defines the state of \mbox{Wi-Fi} node $W^{k}$ in a particular \mbox{LTE-U} network state $S_{l}$. $g(W^{k},S_{l})$ is given as 
\begin{equation}
	\begin{cases} 
		0 & if \ \exists\ L^{i}\ \in EDT(W^{k})\ \land\ \chi_{S_{l}}^{i}=1\\
		1 & else
	\end{cases}
\end{equation}
The function $g(W^{k},S_{l})$ assigns $0$ to a \mbox{Wi-Fi} node if there exists any \mbox{LTE-U} node transmitting in its region indicating those \mbox{Wi-Fi} nodes will be inactive in the current \mbox{LTE-U} network state. $g(W^{k},S_{l})$ assigns 1 to $W^{k}$ if no LTE-U node is transmitting in the EDT region of $W^{k}$, indicating they can take part in transmission in current LTE-U network state. This function helps in determining the current CSMA graph in LTE-U network state $S_{l}$.
\newline 
Let $W_{a}(S_{l})$ denote the set of \mbox{Wi-Fi} nodes which can transmit in LTE-U network state $S_{l}$ and $|W_{a}(S_{l})|=N_{W_{a}}$. $W_{a}(S_{l})$ is then used to construct the CSMA graph of the \mbox{Wi-Fi} network. The nodes with $g()$ output as 1 will be involved in construction of the CSMA graph. The set $W_{a}(S_{l})$ is identified as follows: 
\begin{equation}
W_{a}(S_{l})=\{\ W^{k},\ k\in (1,N_{W})\ |\ g(W^{k},S_{l})=1\ \}
\end{equation}
According to BOE model, independent states set is the set which contains all possible Wi-Fi network states. A Wi-Fi network state is valid if no two \mbox{Wi-Fi} nodes within CSMA range are in state~1 simultaneously. The independent states set in LTE-U network state $S_{l}$ \emph{i.e.,} $IS(S_{l})$  can be built as~follows.\\
Let $Q_{m}(S_{l})$ which is of size $N_{W_{a}}$ represent $m^{th}$ independent state of the \mbox{Wi-Fi} network in \mbox{LTE-U} network state $S_{l}$ and $Q_{m}^{l}(S_{l})$ is the state of $l^{th}$ \mbox{Wi-Fi} node in $m^{th}$ independent state. $Q_{m}^{l}(S_{l})$ with 0 represents back-off due to another \mbox{Wi-Fi} transmission whereas 1 represents on-going transmission. $Q_{m}(S_{l})$ is given as
\begin{eqnarray}
Q_{m}(S_{l})&=&\{\ Q_{m}^{l}(S_{l}),\ l \in (1,N_{W_{a}})\ \}\ |\  Q_{m}^{l}(S_{l})=  1\ \nonumber \\  & & if\ \forall\ W^{k} \in\ CST(W^{l}),\ Q_{m}^{k}(S_{l})=0 
\end{eqnarray}
The above equation tries to assign 1 to $Q_{m}^{l}(S_{l})$ wherever possible, but 1 is not assigned if 1 is already assigned for one of the \mbox{Wi-Fi} nodes within its CSMA range.
\newline 
$IS(S_{l}) = \{Q_{m}(S_{l}) \}$, where $Q_{m}(S_{l})$ is a feasible independent state.
Maximum Independent States (MIS) are defined as $MIS(S_{l})=\{Q_{m}(S_{l})\ |\ \ \ \sum_{l=1}^{N_{W_{a}}} Q_{m}^{l}(S_{l})\ \ is\ maximized\}$ and $|MIS(S_{l})|\ = N_{MIS}(S_{l})$. $MIS(S_{l})$ tries to identify $Q_{m}(S_{l})$ in which the maximum number of \mbox{Wi-Fi} nodes are simultaneously transmitting.
\newline 
BOE model argues that the average of the MIS computed gives the behavior of Wi-Fi network, $BOE(W_{a}(S_{l}))$ accounts for the normalized throughput distribution of Wi-Fi nodes in set $W_{a}(S_{l})$. $BOE(W_{a}^{i}(S_{l}))$ gives the normalized throughput of the Wi-Fi node $W_{a}^{i}(S_{l})$. $BOE(W_{a}(S_{l}))$ is computed as follows: 
	\begin{equation}
	BOE(W_{a}(S_{l}))=\frac{\sum_{n=1}^{N_{MIS}(S_{l})}MIS^{n}(S_{l})}{N_{MIS}(S_{l})}
	\end{equation}
	We derive normalized throughput distribution of all Wi-Fi nodes as $W_{nor}^{i}(S_{l}), i\in(1,N_{W})$.
	\begin{equation}
W_{nor}^{i}(S_{l}) = 	
\begin{cases} 
BOE(W^{i}) & \text{when  }  g(W^{i},S_{l})=1\\
0 & \text{when  }  g(W^{i},S_{l})=0 
\end{cases}
	\end{equation}
	\mbox{Wi-Fi} nodes, $W - W_{a}(S_{l})$, will be having zero  normalized throughput and normalized throughput of nodes in $W_{a}(S_{l})$ will be correspondingly mapped from $BOE(W_{a}(S_{l}))$ to $W_{nor}(S_{l})$ in \mbox{LTE-U} network state $S_{l}$.
\newline
With the above equations, \mbox{Wi-Fi} network normalized throughput distribution over time and in all LTE-U network states can be computed~as follows:
\begin{equation}
W_{nor}=\sum_{L_{set}}\sum_{t=0}^{T_{frame}}\ \frac{\psi_{s}(t) \times W_{nor}(s) \times \Delta t}{T_{frame}}
\end{equation}
\newline 
where, $L_{set}$ is the set of feasible \mbox{LTE-U} network states with a non-zero probability at some point of time. $L_{set}$ is constructed during computations of $CCD$, $AC$, and $AT$. As we defined our model to be continuous over time, we can replace summation over time to integration to convey the notion of continuity.
\begin{equation}
W_{nor}=\sum_{L_{set}}\int_{t=0}^{Tframe} \frac{\psi_{s}(t) \times W_{nor}(s)\ dt}{T_{frame}}
\label{intint}
\end{equation}
\newline 
Eqn.~(\ref{intint}) represents the normalized throughput distribution of all Wi-Fi nodes in the network. Actual throughput can be computed by multiplying normalized throughput distribution with single link throughput as followed by BOE model~\cite{boe}. Wi-Fi throughput equation when all Wi-Fi nodes fall under same CSMA region is derived by Bianchi model~\cite{bianchi} using stochastic analysis as follows:
\begin{equation}
Z=\frac{P_{s}P_{tr}E[P]}{(1-P_{tr})\sigma+P_{tr}P_{s}T_{s}+P_{tr}(1-P_{s})T_{c}}
\label{eqn18}
\end{equation}
\color{black}
where, $P_{tr}$ denotes the probability that at least one Wi-Fi node is transmitting, $P_{s}$ is the probability that a transmission is successful, $\sigma$ is the time slot considered in CSMA/CA protocol, $E[P]$ is the packet payload, $T_{s}$ is the time taken for successful transmission, and $T_{c}$ is the time spent in collision. Single link throughput is defined as the throughput a node gets under saturated buffer conditions and it is only the channel contender in the network, which is obtained by substituting corresponding $P_{S}$ and $P_{tr}$ values in Eqn.~(\ref{eqn18}). $Z$ denotes the Wi-Fi throughput for any number of Wi-Fi nodes in Eqn.~(\ref{eqn18}), let $Z_{1}$ denote the single link Wi-Fi throughput then actual Wi-Fi throughput in the LTE-U Wi-Fi coexistent network is derived~as 
\begin{equation}
W_{Thr}=W_{nor} \times Z_{1};
\end{equation}
\newline 
We define ``air time fraction" ($A$) to study the coexistence between \mbox{Wi-Fi} and \mbox{LTE-U} networks in spatial distribution scenarios. Air time fraction is defined as the fraction of time a node gets to access the channel \emph{i.e.,} it includes both successful transmissions and collisions. Collisions are included in air time as we are studying coexistence in terms of channel access opportunity. \mbox{LTE-U} node air time fraction ($A_{L_{i}}$) is straightforward from its duty cycle: $A_{L}^{i}=D^{i}$. But for \mbox{Wi-Fi} case, the total time allotted for \mbox{Wi-Fi} is spent in transmissions, collisions, and wasted by being idle. So to get original air time, we have to take CSMA factor ($\Omega$) into consideration. $\Omega$ can be defined as the probability for which the considered slot time will be having either a collision or a successful transmission. From~\cite{bianchi}, we get
\begin{equation} 
\Omega=\frac{P_{tr}P_{s}T_{s}+P_{tr}(1-P_{s})T_{c}}{(1-P_{tr})\sigma+P_{tr}P_{s}T_{s}+P_{tr}(1-P_{s})T_{c}}
\end{equation}
Wi-Fi node air time fraction ($A_{W}^{i}$) is given as follows:
\begin{equation}
A_{W}^{i}=W_{nor}^{i} \times \Omega_{1}
\end{equation}
\newline 
where, $\Omega_{1}$ is the value of $\Omega$ evaluated when only one Wi-Fi node is transmitting in the network. The above proposed equations to calculate air times are also validated later using simulation studies in Section~\ref{PE}.

\section{Performance Evaluation}\label{PE}
To validate the proposed model and study the coexistence between LTE-U and Wi-Fi, we developed a system level simulator for Wi-Fi--LTE-U networks using MATLAB. We first validate our MATLAB simulator with the mathematical models of \mbox{Wi-Fi--Wi-Fi}\cite{bianchi} and \mbox{Wi-Fi}--\mbox{LTE-U}\cite{WL} nodes where all nodes can sense presence of all others. Further, the available model for \mbox{Wi-Fi--Wi-Fi}\cite{boe} for spatially distributed Wi-Fi nodes is also used to validate the MATLAB simulator. Then we use the simulator to validate the proposed analytical model for spatially distributed \mbox{Wi-Fi--LTE-U} networks. The system parameters are shown in~Table~\ref{Parameter1}.
\begin{figure}[htb!]
	\vspace{-0.3cm}
	\begin{subfigure}{0.5\linewidth}
		\includegraphics[width=1.0\linewidth]{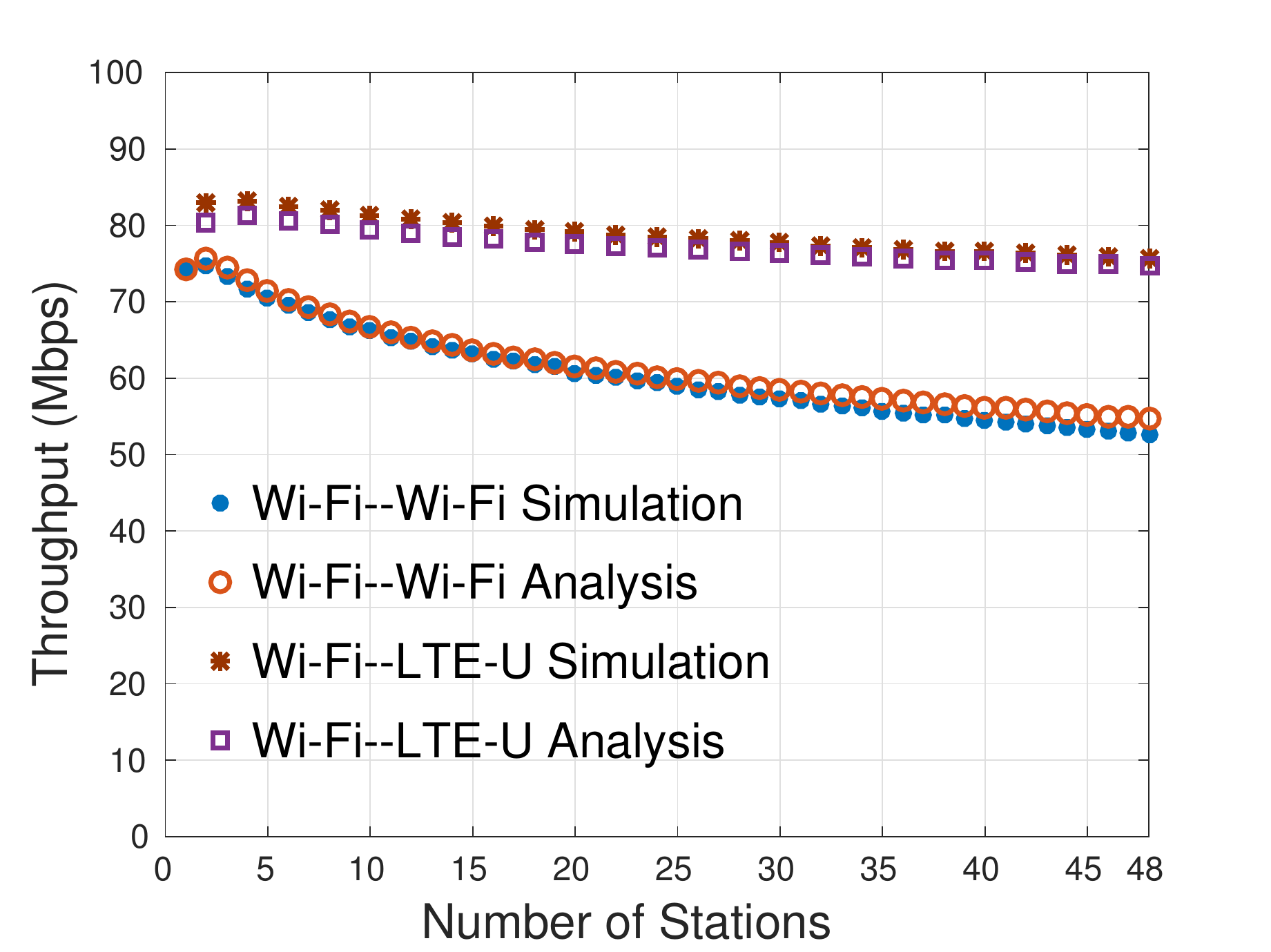}
		\caption{\mbox{Wi-Fi} and LTE-U model.}
		\label{wifiltu}
	\end{subfigure}
	\hspace{-0.3cm}
	\begin{subfigure}{0.5\linewidth}
		\includegraphics[width=0.95\linewidth]{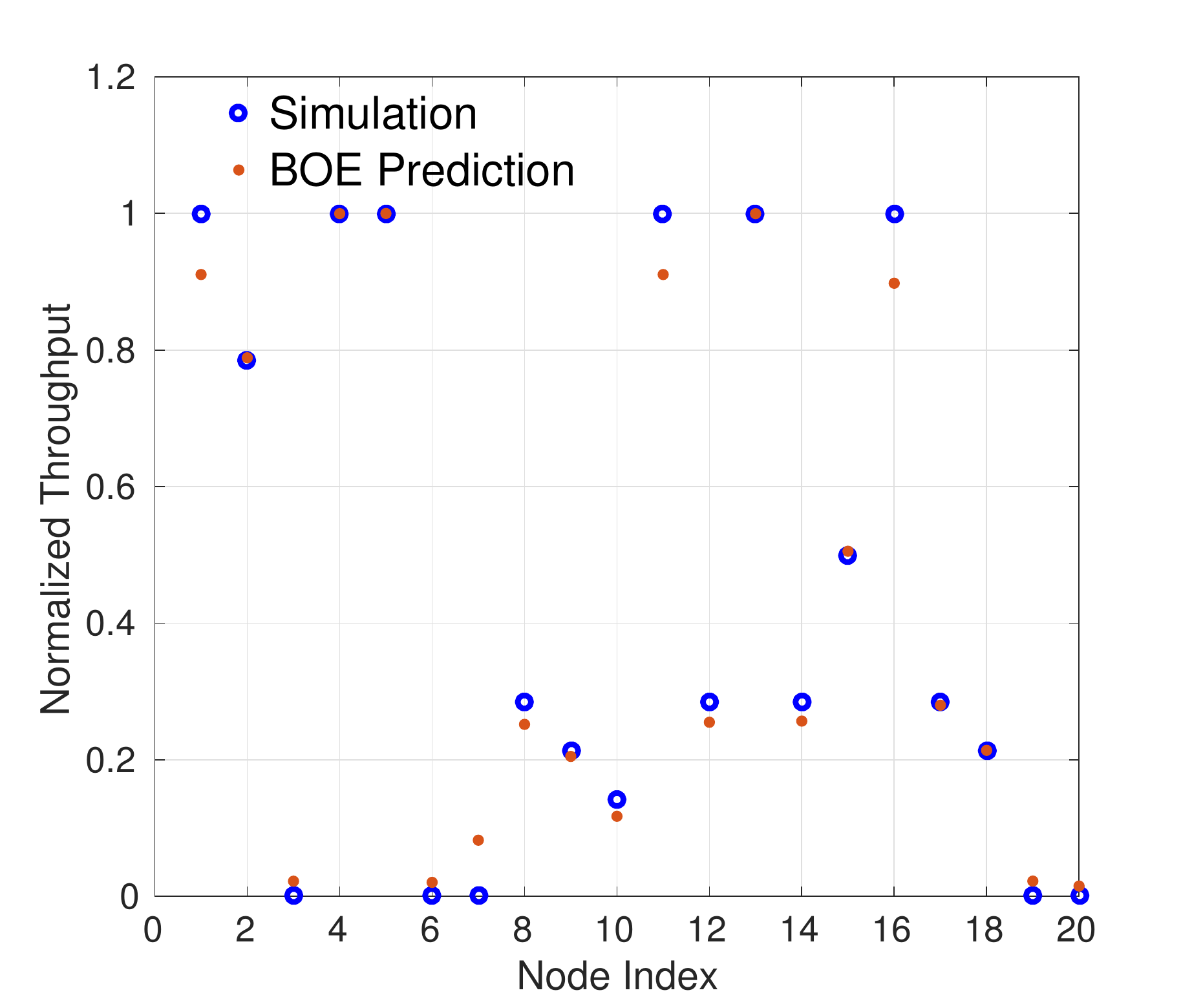}
		\caption{BoE Model.}
		\label{boe}
	\end{subfigure}
	\caption{Validation of Simulator with the help of existing analytical models~\cite{bianchi},\cite{boe},\cite{WL}.}
	\vspace{-0.4cm}
\end{figure}
\begin{table}[htb!]
	\caption{System Parameters}
	\centering
	\begin{tabular}{|p{3.1cm}| p{4.7cm}|}
		\hline\hline
		\bfseries
		\  Parameter&\bfseries Value \\ [0.2ex]
		\hline
		$CW_{min}$  & 16 \\
		\hline
		$CW_{max}$  & 1024 \\
		\hline

		PHY Header Size & 128 bits \\
		\hline
			MAC Header Size & 272 bits \\
		\hline
		ACK Size & 240 bits \\
			\hline
			Maximum PDU & 4 \\
			\hline
		Payload Size & 8148 bits \\
		\hline
		Slot Time & 9 $\mu$sec\\
		\hline
		DIFS & 34 $\mu$sec \\
		\hline
		SIFS & 16 $\mu$sec \\
		\hline
			Max-retry limit & 6 \\
		\hline
		EDT value & -62 dBm \\
		\hline
				CST value  & -82 dBm \\
				\hline

		Tx Power of AP and eNB & 20 dBm \\
		\hline
			Operating Freq. & 5.3 GHz \\
		\hline
		System Noise & -101 dBm \\
		\hline
		\mbox{\mbox{Wi-Fi}} PHY Rate & 130 Mbps \\
		\hline
		\mbox{\mbox{Wi-Fi}} ACK Rate & 26 Mbps\\
		\hline
		\mbox{\mbox{Wi-Fi}} Header Rates &  6.5Mbps \\
		\hline
		\mbox{LTE-U} PHY Rate & 93.24 Mbps \\
		\hline
		Traffic & Full buffer via saturated UDP flows \\
		\hline
			Channel & No shadow/Rayleigh fading \\
		\hline
		Path Loss Model\cite{PL}   & 36.7log10(d[m])+22.7+26log10(freq[GHz])  \\
		\hline
		 $T_{frame}$ & 40 msec \\
		\hline
		Duty cycle of $i^{th}$ LTE-U node ($D_{L}^{i}$) & min(0.95, 1/(1+Number of nodes inside EDT of $i^{th}$ LTE-U node))\\
		\hline
		Simulation Time & 50 sec \\
		\hline
	\end{tabular}
	\label{Parameter1}
\end{table}
\begin{table*}[htb!]
	\caption{Comparison of Analytical and Simulation based throughputs (in Mbps) of Wi-Fi--LTE-U networks for Topologies 1 - 3} 
	\centering
	\begin{tabular}{|c|>{\columncolor[gray]{0.8}}c|c|>{\columncolor[gray]{0.8}}c|c|>{\columncolor[gray]{0.8}}c|c|>{\columncolor[gray]{0.8}}c|c|c|}
		\hline
		\\[-1em]
		\bfseries \hspace{-0.25cm} Node Index \hspace{-0.25cm} &\multicolumn{3}{|c|}{\bfseries Topology \#1}&\multicolumn{3}{|c|}{\bfseries Topology \#2}& \multicolumn{3}{|c|}{\bfseries Topology \#3}  \\\hline
		\\[-1em]
		- & Analysis &Simulation &Error (\%) &Analysis &Simulation &Error (\%) &Analysis &Simulation &Error (\%) \\\hline
		\\[-1em]
		W1 & 31.51 & 31.31 &0.63\% & 34.60& 33.25 &3.90\% &  55.61 & 55.13 &0.87\%  \\\hline
		\\[-1em]
		W2 & 35.22 & 35.24 &0.05\% & 51.25 & 50.52 &1.44\%  &30.89 & 31.38 &1.56 \% \\\hline
		\\[-1em]
		W3 & 42.02 & 41.28 & 1.79\% & 6.83 & 7.25 &5.79\% &24.71 & 24.09 &2.57\% \\\hline
		\\[-1em]
		W4 & 48.20& 47.23  &2.05\% & 59.32 & 58.44 &1.50\% &  - & - & -  \\\hline
		\\[-1em]
		W5 & 33.37 & 32.93 &1.33\% &59.32 & 58.46 &1.47\% & - & - & - \\\hline
		\\[-1em]
		L1 & 18.64 & 18.16 &2.6\% & 23.31 & 23.20 &0.47\% & 23.31 & 22.94 &1.61\%  \\\hline
		L2 & 23.31 & 22.81 &2.19\% & 23.31 & 22.91 & 1.74\% & 31.08 & 31.07&0.03 \% \\\hline
		L3 & 31.08 & 30.92 &0.51\% & 23.31 & 23.17 & 0.60\% & 31.08 & 30.90 &0.58 \%\\\hline
		L4 & 31.08 & 30.79 &0.94\% & 18.64 & 18.22 &2.30\% &  - & - & - \\\hline
		L5 & 31.08 & 30.92 &0.51\% &31.08 & 30.98 &0.32\% &  - & - & - \\\hline
	\end{tabular}\label{T3}
\end{table*}
\begin{figure}[htb!]
	\begin{center}
		\includegraphics[width=8cm,height=3cm]{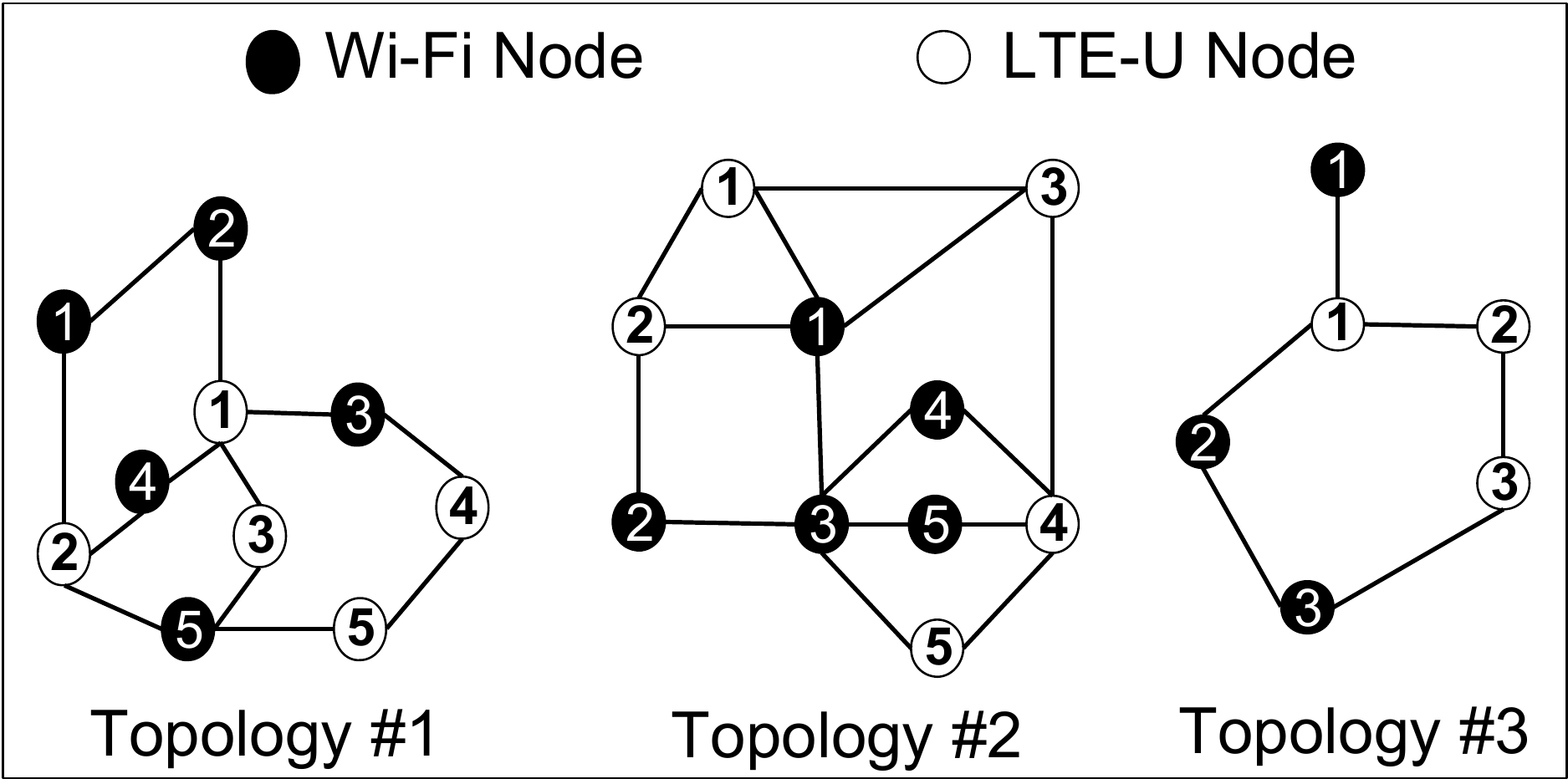}
		\caption{Experimental topologies.}
		\label{scenarios}
	\end{center}
	\vspace{-0.5cm}
\end{figure}
\begin{figure*}[htb!]
	\begin{subfigure}{0.33\linewidth}
		\includegraphics[width=1.05\linewidth]{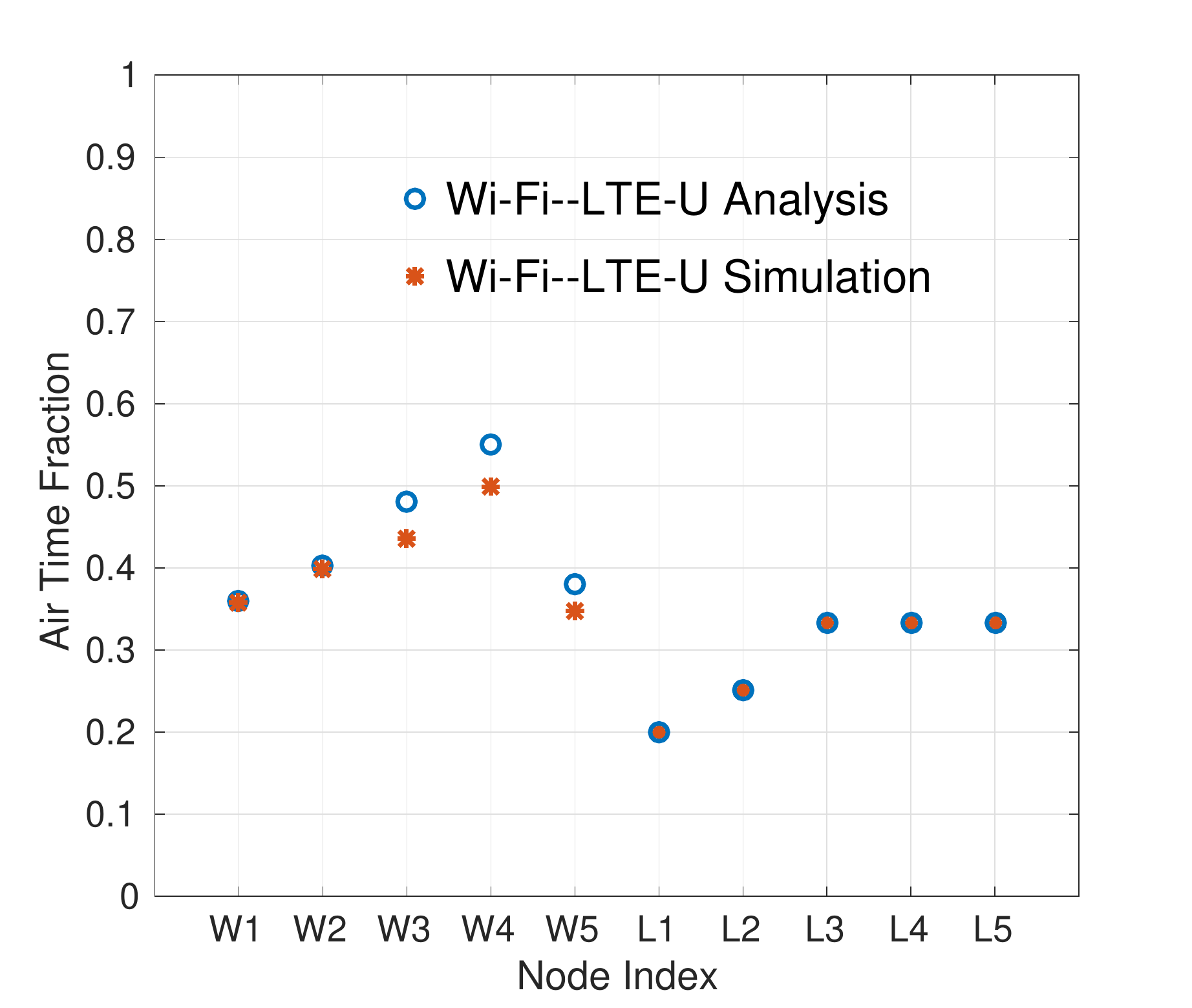}
		\caption{Topology \#1.}
		\label{airtime1}
	\end{subfigure}
	\begin{subfigure}{0.33\linewidth}
		\includegraphics[width=1.05\linewidth]{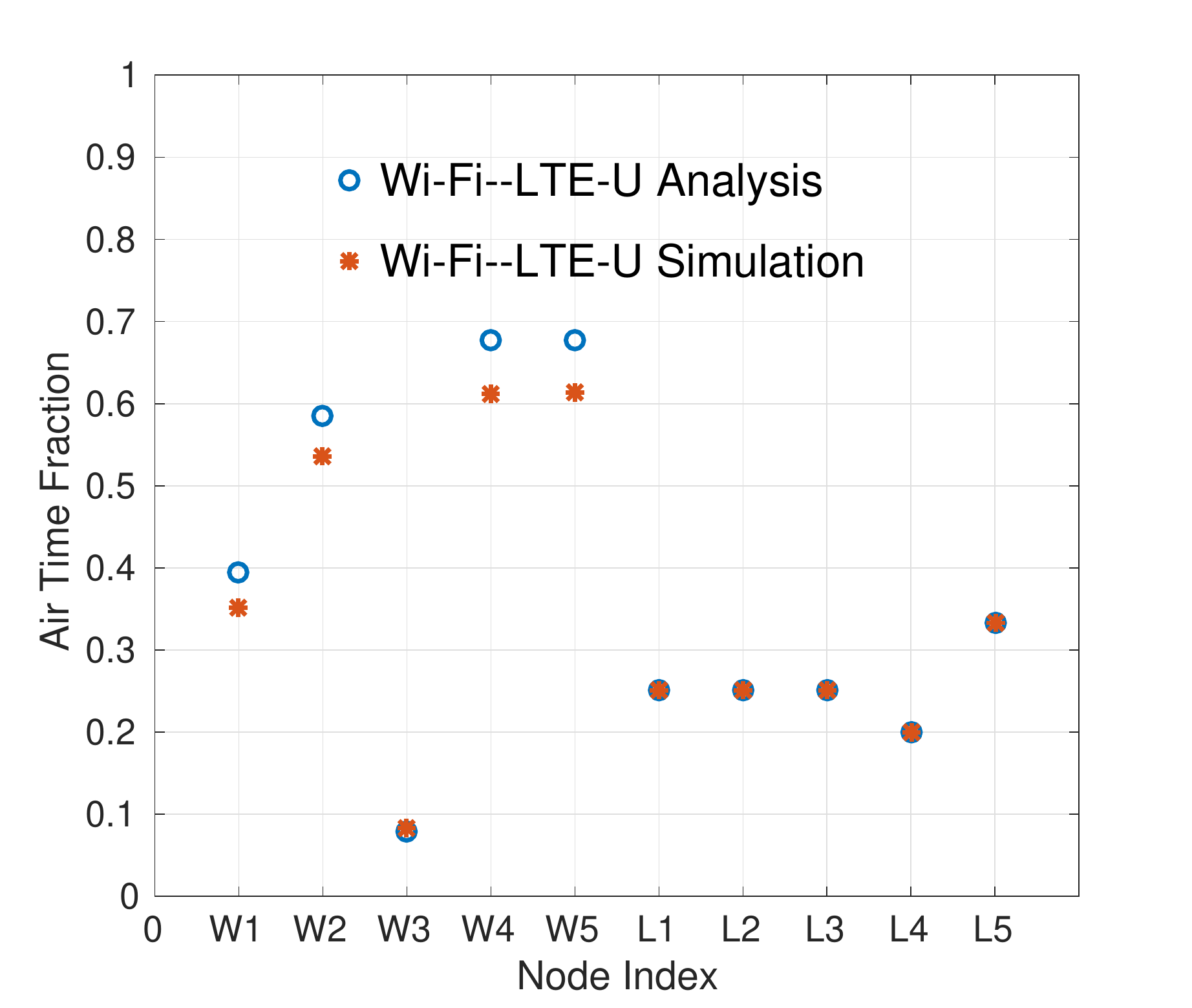}
		\caption{Topology \#2.}
		\label{airtime2}
	\end{subfigure}
	\begin{subfigure}{0.33\linewidth}
		\includegraphics[width=1.05\linewidth]{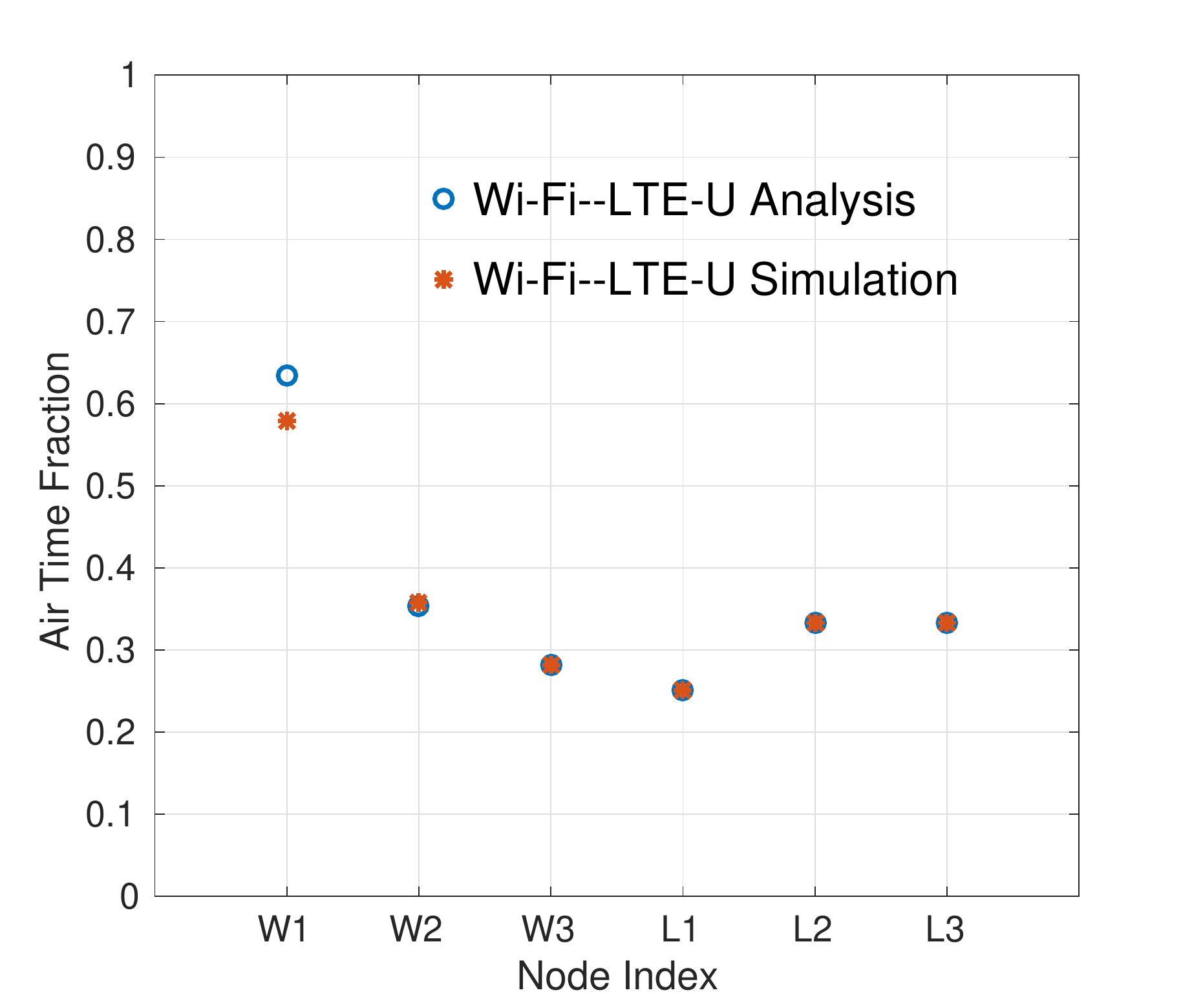}
		\caption{Topology \#3.}
		\label{airtime3}
	\end{subfigure}
	\caption{Comparison of air time fraction obtained through simulation and analytical studies.}
	\label{AirTimes}
\end{figure*}
\begin{figure*}[t]
	\vspace{-0.3cm}

	\begin{subfigure}{0.5\linewidth}
		\includegraphics[width=0.9\linewidth]{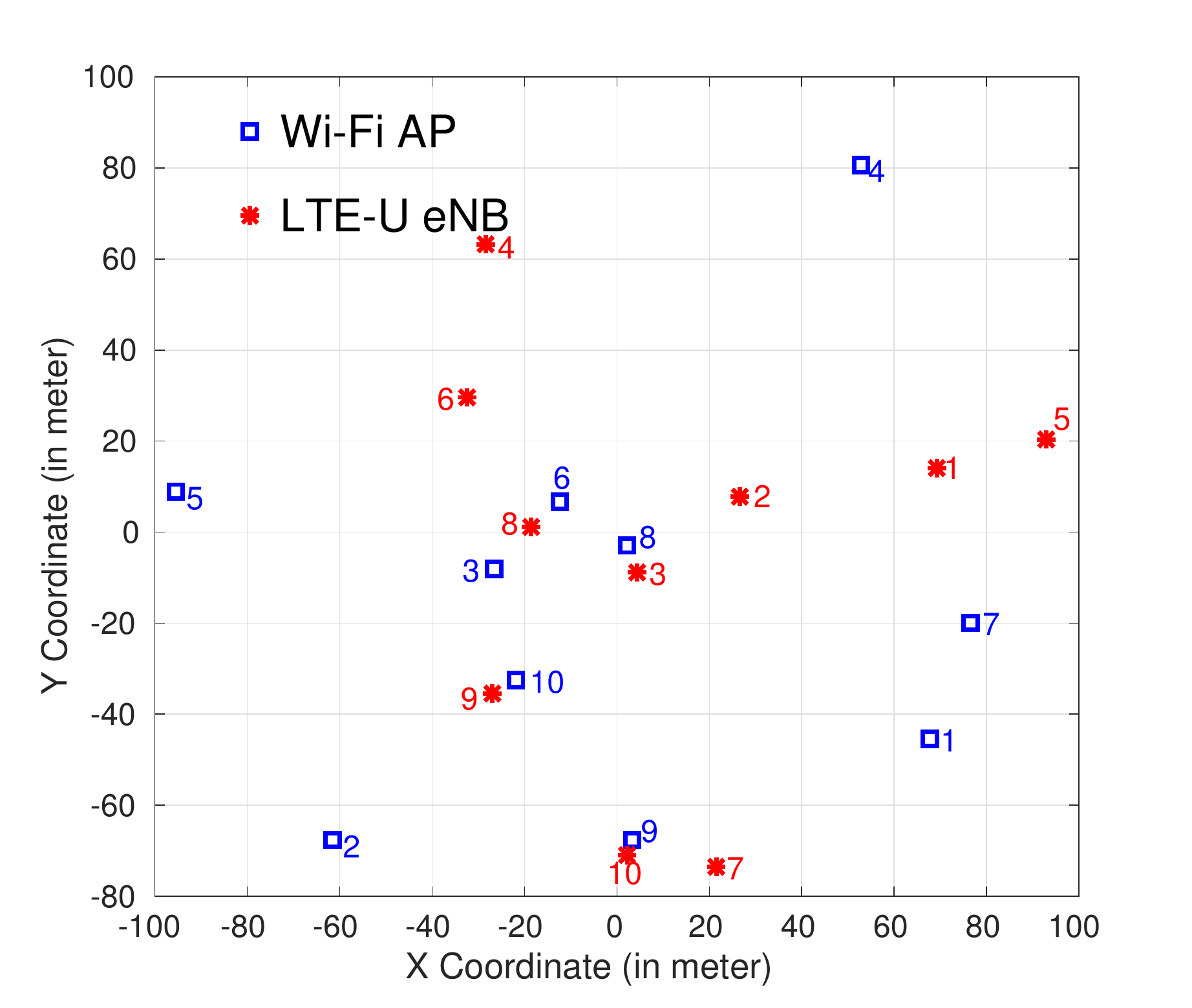}
		\caption{Random placement generated.}
		\label{placement}
	\end{subfigure}
	\begin{subfigure}{0.5\linewidth}

		\includegraphics[width=0.9\linewidth]{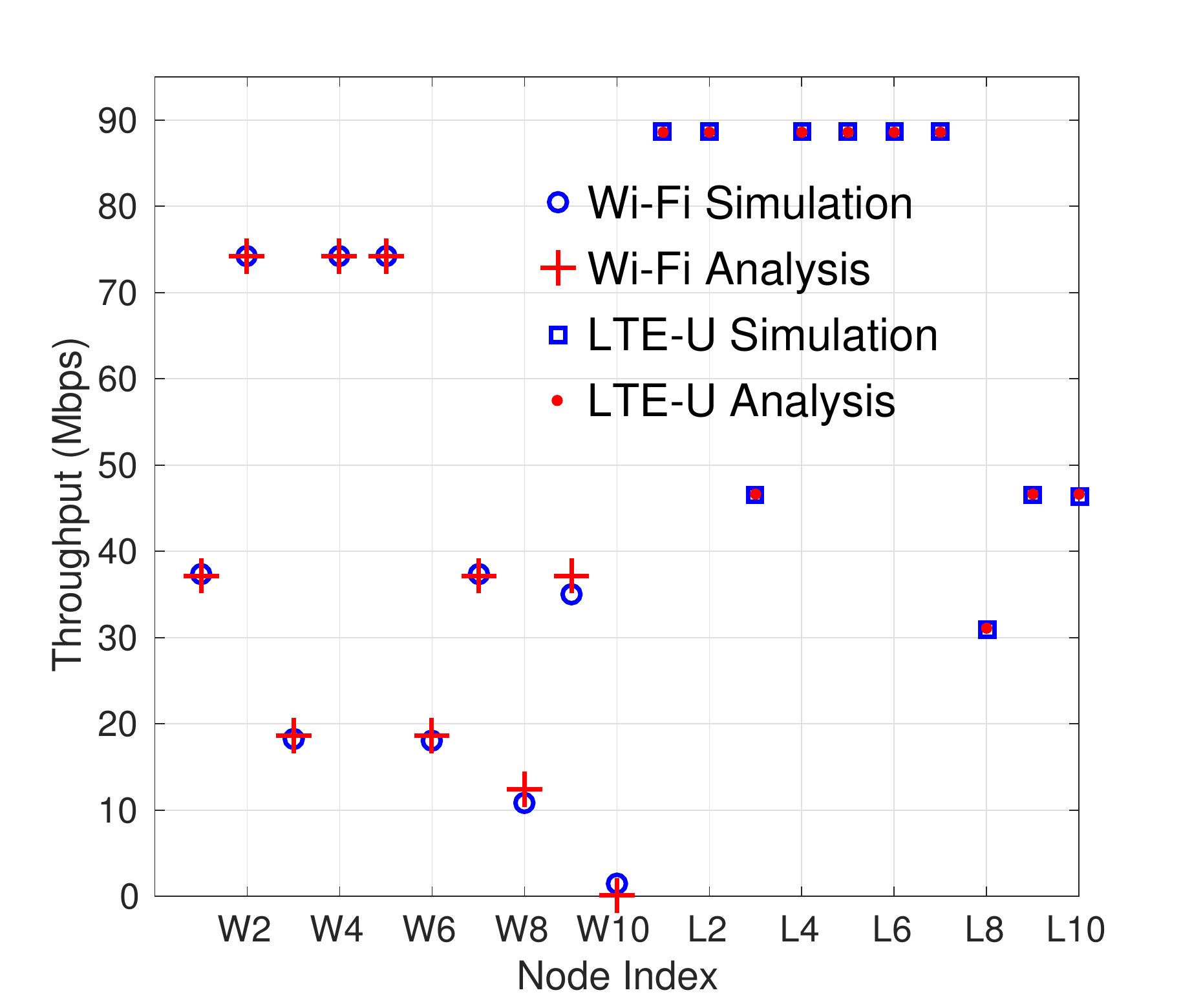}
		\caption{Model validation.}
		\label{modelvalidation}
	\end{subfigure}
	\hspace{-0.3cm}
	\caption{One of the randomly generated topologies and the corresponding throughput result of each node obtained through simulation and analytical studies.}
	\vspace{-0.4cm}
\end{figure*}

\subsection{Simulator Validation}
Fig.~\ref{wifiltu} shows a comparison of \mbox{Wi-Fi} analysis presented in~\cite{bianchi} and \mbox{Wi-Fi}--\mbox{LTE-U} analysis presented in~\cite{WL} with our MATLAB simulator. In Fig.~\ref{wifiltu}, throughput is plotted with respect to the number of stations in the network. Fig.~\ref{boe} shows the comparison graph of BoE model\cite{boe} with the results from our simulator. The percentage error in the throughput computation of Wi-Fi--Wi-Fi network (within CSMA range) is 1.91\% while it is 1.92\% for Wi-Fi--LTE-U (within EDT Range). In Fig.~\ref{boe}, normalized throughput for each node is shown for 20-node topology.  The mean normalized link/node throughput error (metric defined by BOE~\cite{boe}) in Wi-Fi--Wi-Fi network (spatially distributed and partially overlapping ) is observed to be 2\% for 20-node topology. We can see that the values of the existing models and our simulator are closely matching which proves the accuracy of simulator developed in simulating Wi-Fi--LTE-U network scenarios. We have used the same simulator to validate our proposed analytical model in the next section.
\subsection{Validation of the proposed model}
In this section, we have validated our analytical model for three specific topologies as shown in Fig.~\ref{scenarios}. The total number of nodes are 10, 10, and 6 in topology 1, 2, and 3, respectively. In each topology, we have 50\% \mbox{LTE-U} and 50\% \mbox{Wi-Fi} nodes.  Table~\ref{T3} shows the comparison of simulation and analytical throughputs of each node in all the three topologies. Fig.~\ref{AirTimes} shows the air time prediction using our model for all the three topologies. In Fig.~\ref{AirTimes}, the node index on X-axis represents the index of each node where $Wi$ is $i^{th}$ \mbox{Wi-Fi} node and $Li$ is $i^{th}$ \mbox{LTE-U} node. It can be seen from Table~\ref{T3} that the analytical throughputs of \mbox{LTE-U} and \mbox{Wi-Fi} nodes are closely matching with the simulation throughputs, thereby validating the proposed model. Further, Fig.~\ref{AirTimes} shows the air time fraction each node is getting through simulation and analytical studies.

\subsection{Validation of the proposed model in random topologies}
In the next phase of our validation, we validate our model against multiple randomly generated network topologies. Fig.~\ref{placement} shows a placement of one such randomly generated network topology which consists of 10 \mbox{Wi-Fi} and 10 \mbox{LTE-U} nodes in the area of 200 meter X 200 meter. The same topology is simulated and validated using the proposed analytical model. Fig.~\ref{modelvalidation} shows the simulation and analytical throughput results for each node. From Fig.~\ref{modelvalidation}, we can see that the proposed analytical model closely matches with the simulation results; thereby demonstrating the correctness of the proposed analytical model.
Further, Table~\ref{LinkErr} shows the mean node throughput error associated with the \mbox{Wi-Fi} throughput, \mbox{LTE-U} throughput, and system throughput. The number of nodes in the system are varied from 10 to 40 with a step size of 10. Out of the total number of nodes, 50\% nodes are Wi-Fi and remaining 50\% nodes are LTE-U. Mean throughput error is computed over 50 randomly generated network topologies. We can see that the throughput error is quite low for practical networks topologies. \mbox{LTE-U} throughput can be estimated with greater accuracy because of simple channel access mechanism followed by \mbox{LTE-U}. The reason for observed error is due to the fact of neglected collisions which is observed to be quite less. Further, in the proposed model we have made use of BOE model~\cite{boe} for throughput computation of \mbox{Wi-Fi}--\mbox{Wi-Fi} contention graph. As the BoE model does not consider \mbox{Wi-Fi}--\mbox{Wi-Fi} collisions, one can use a more accurate model from  available \mbox{Wi-Fi}--\mbox{Wi-Fi} models\cite{W2},\cite{W3} to get more accurate results.
For \mbox{Wi-Fi}--\mbox{LTE-U} collisions during each transition, we considered the fact that when $T_{frame}$ is considerably long, number of successful transmissions are dominant over collisions. This point is studied in detail in the next section.
\begin{table}[htb!]
	\centering
	\caption{Mean Throughput Error between Simulation and Analytical results for random topologies.}
	\begin{tabular}{|c|c|c|c|c|}
		\hline
		\bfseries \# Nodes &\bfseries 10-node&\bfseries 20-node&\bfseries 30-node&\bfseries 40-node\\
		\hline
		$WiFi_{Thr}$ & 0.49\% & 0.95\% & 1.61\% & 2.27\% \\ \hline
		$LTE-U_{Thr}$& 0.01\% & 0.01\%& 0.02\% & 0.02\%\\ \hline
		$System_{Thr}$& 0.25\% & 0.48\%& 0.81\% & 1.14\%\\ \hline
	\end{tabular}\label{LinkErr}
\end{table}
\subsection{Effect of Variation of $T_{frame}$ on Throughput}
\begin{figure}[htb!]
	\begin{center}
		\includegraphics[width=0.95\linewidth]{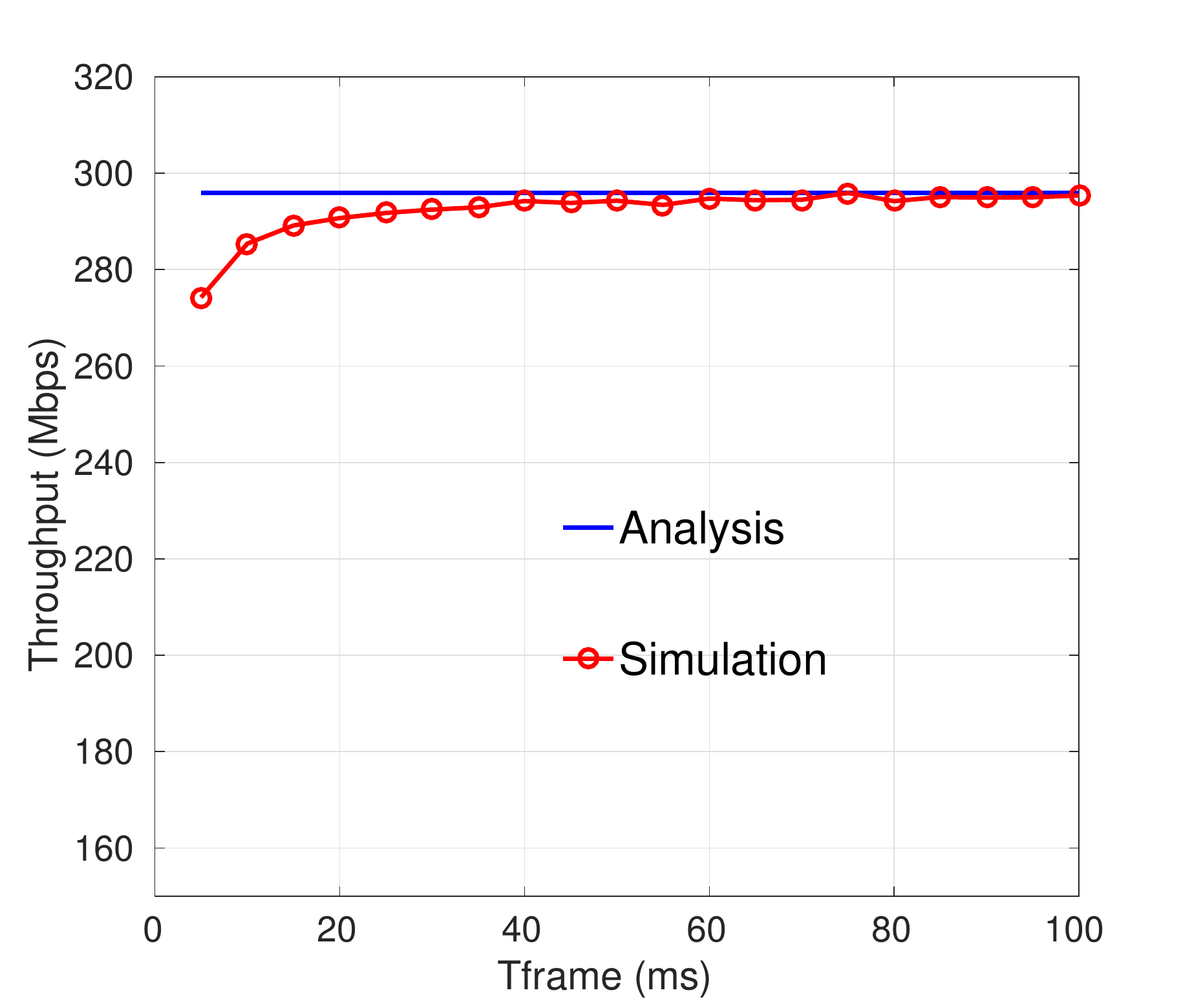}
		\caption{System throughput vs duration of $T_{frame}$.}
		\label{Tframe}
	\end{center}
	\vspace{-0.7cm}
\end{figure}

During \mbox{LTE-U} state transitions, collisions happen between \mbox{Wi-Fi} and \mbox{LTE-U} due to lack of coordination between \mbox{Wi-Fi} and \mbox{LTE-U}. \mbox{LTE-U} does not have any backoff mechanism to fallback if the medium is busy. This leads to collisions which results in channel wastage and doubling of Wi-Fi contention window. In our modeling, we did not consider the occurrence of these events. Note that for longer $T_{frame}$, number of successful transmission events significantly dominates the collision events. Under those circumstances, collisions really have negligible impact on determining the system throughput. We have performed simulations to validate this claim. Results for one such random topology is shown in Fig.~\ref{Tframe}. These results are also in accordance with our argument given earlier.
\begin{figure*}[htb!]
	\vspace{-0.3cm}
	\begin{subfigure}{0.33\linewidth}
		\includegraphics[width=1.05\linewidth]{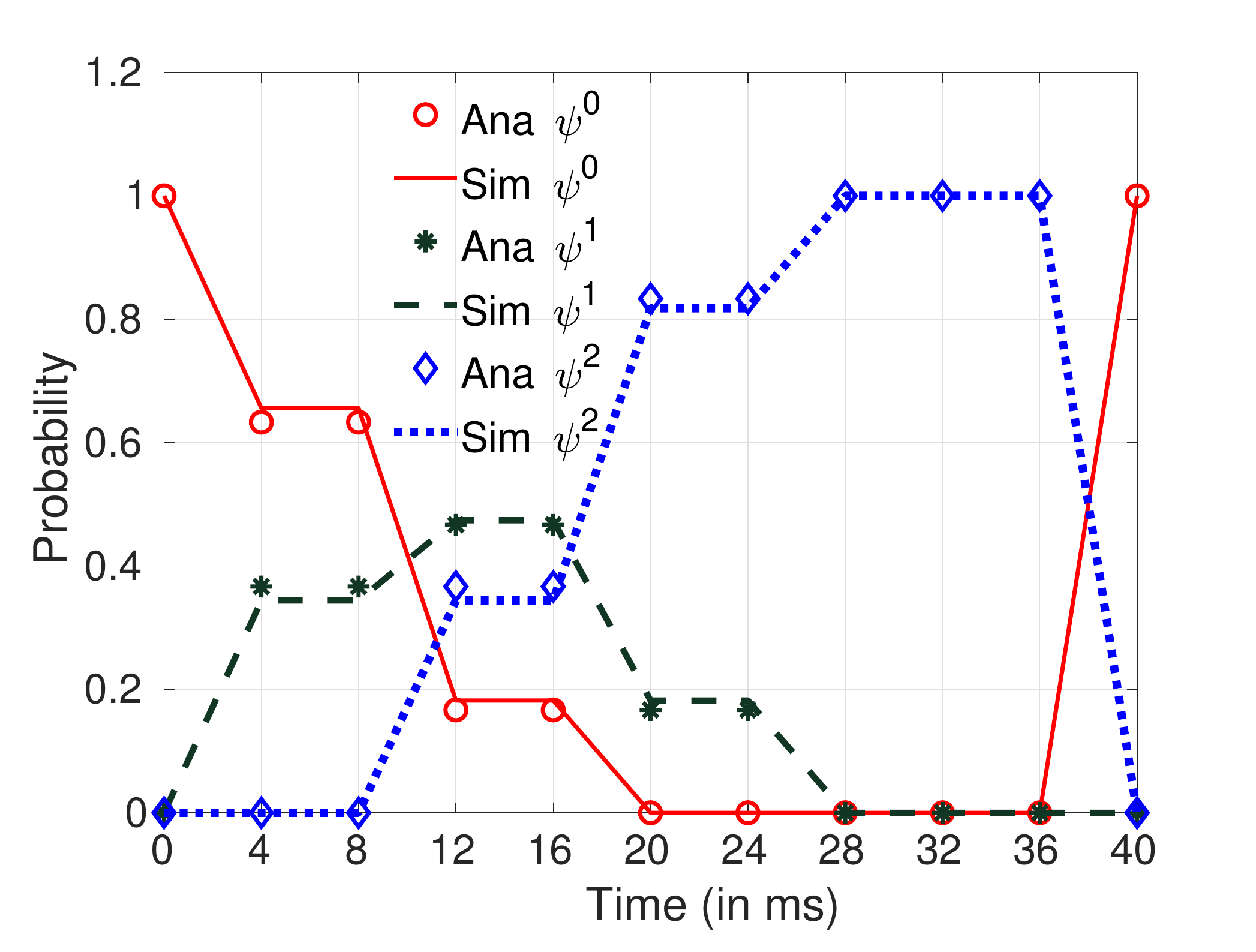}
		\caption{State Probability ($\psi^{\{0,1,2\}}$) of node L1.}
		\label{L1}
	\end{subfigure}
	\hspace{-0.3cm}
	\begin{subfigure}{0.33\linewidth}
		\includegraphics[width=1.05\linewidth]{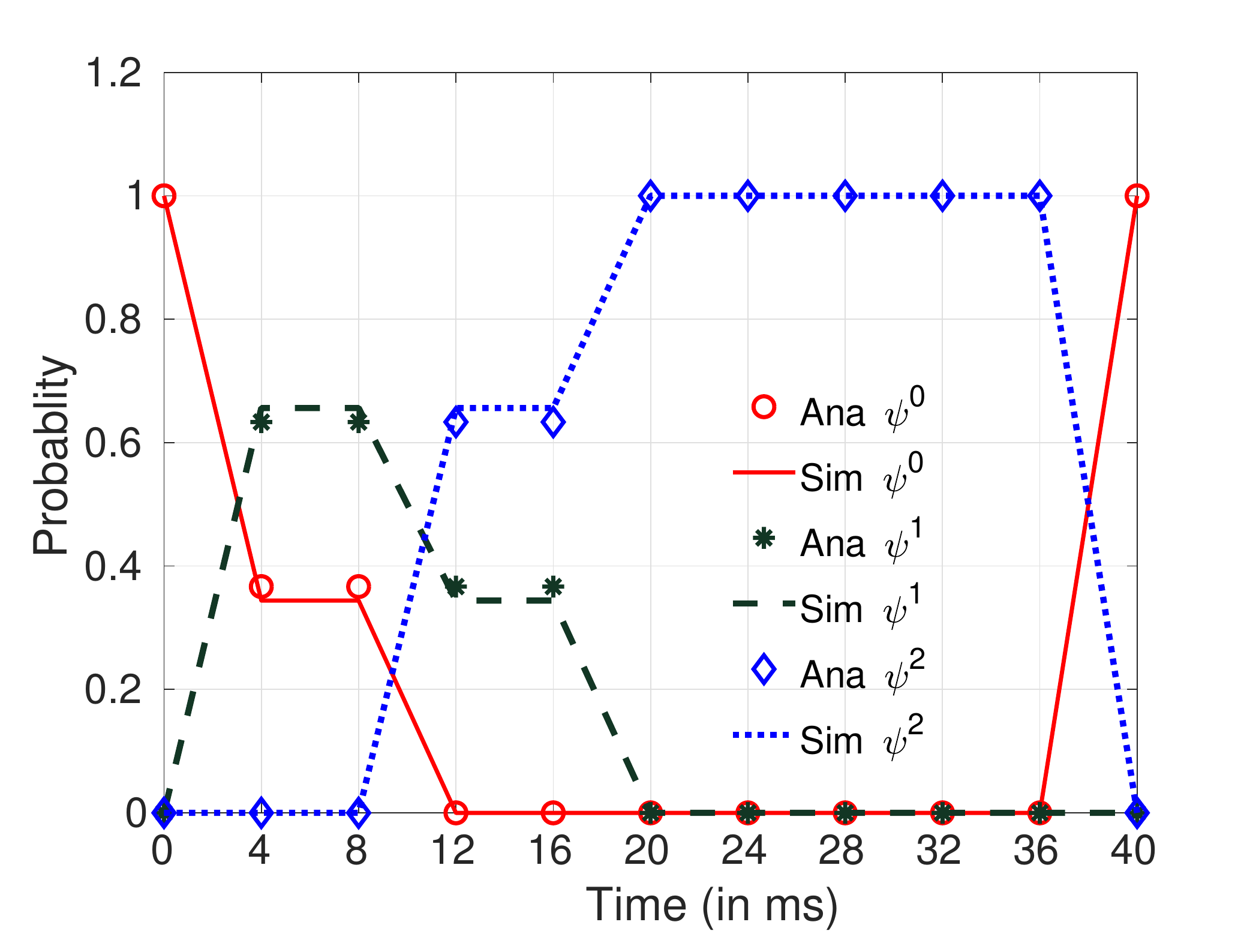}
		\caption{State Probability ($\psi^{\{0,1,2\}}$) of node L2.}
		\label{L2}
	\end{subfigure}
	\begin{subfigure}{0.33\linewidth}
		\includegraphics[width=1.05\linewidth]{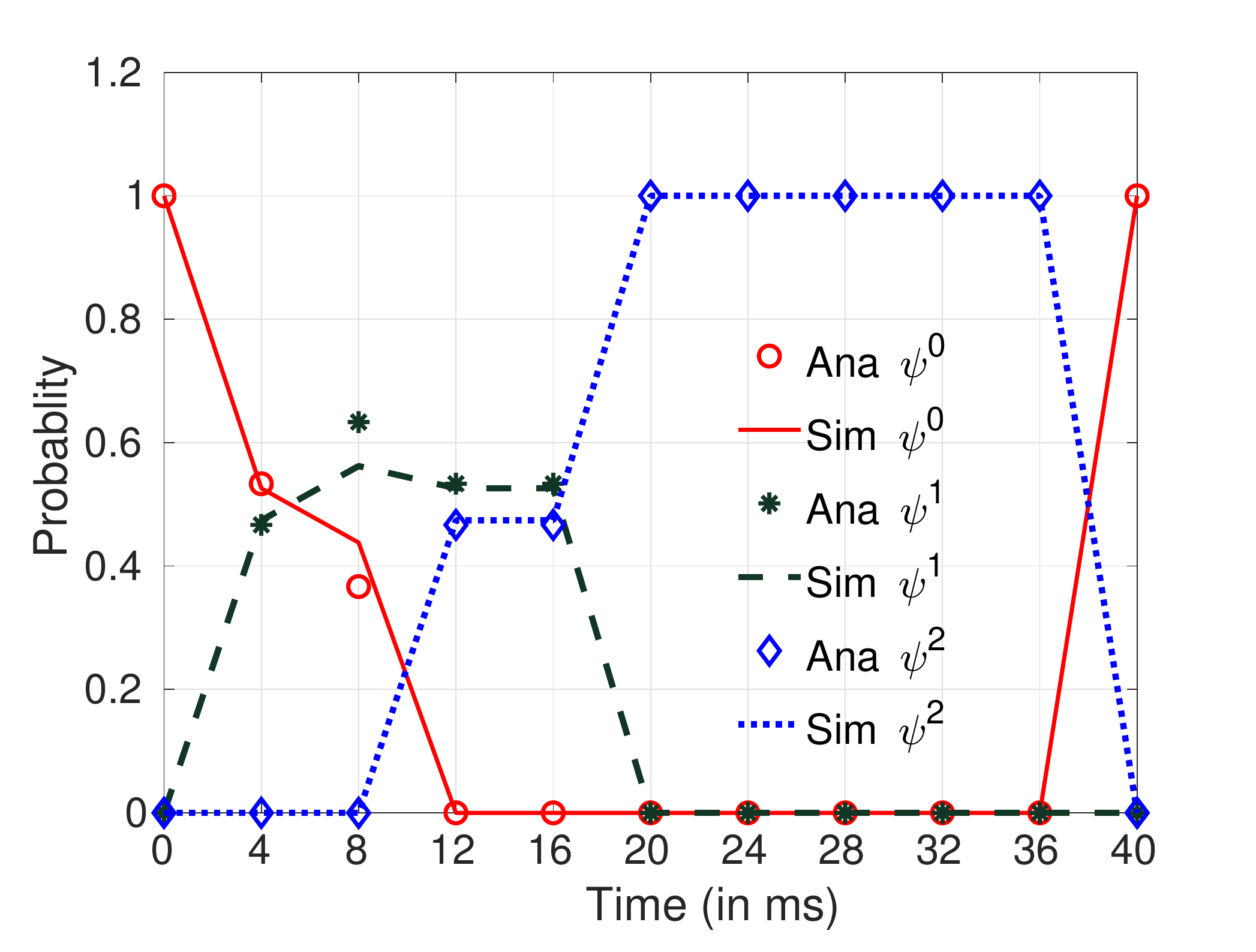}
		\caption{State Probability ($\psi^{\{0,1,2\}}$) of node L3.}
		\label{L3}
	\end{subfigure}
	\begin{subfigure}{0.49\linewidth}
		\includegraphics[width=1.05\linewidth]{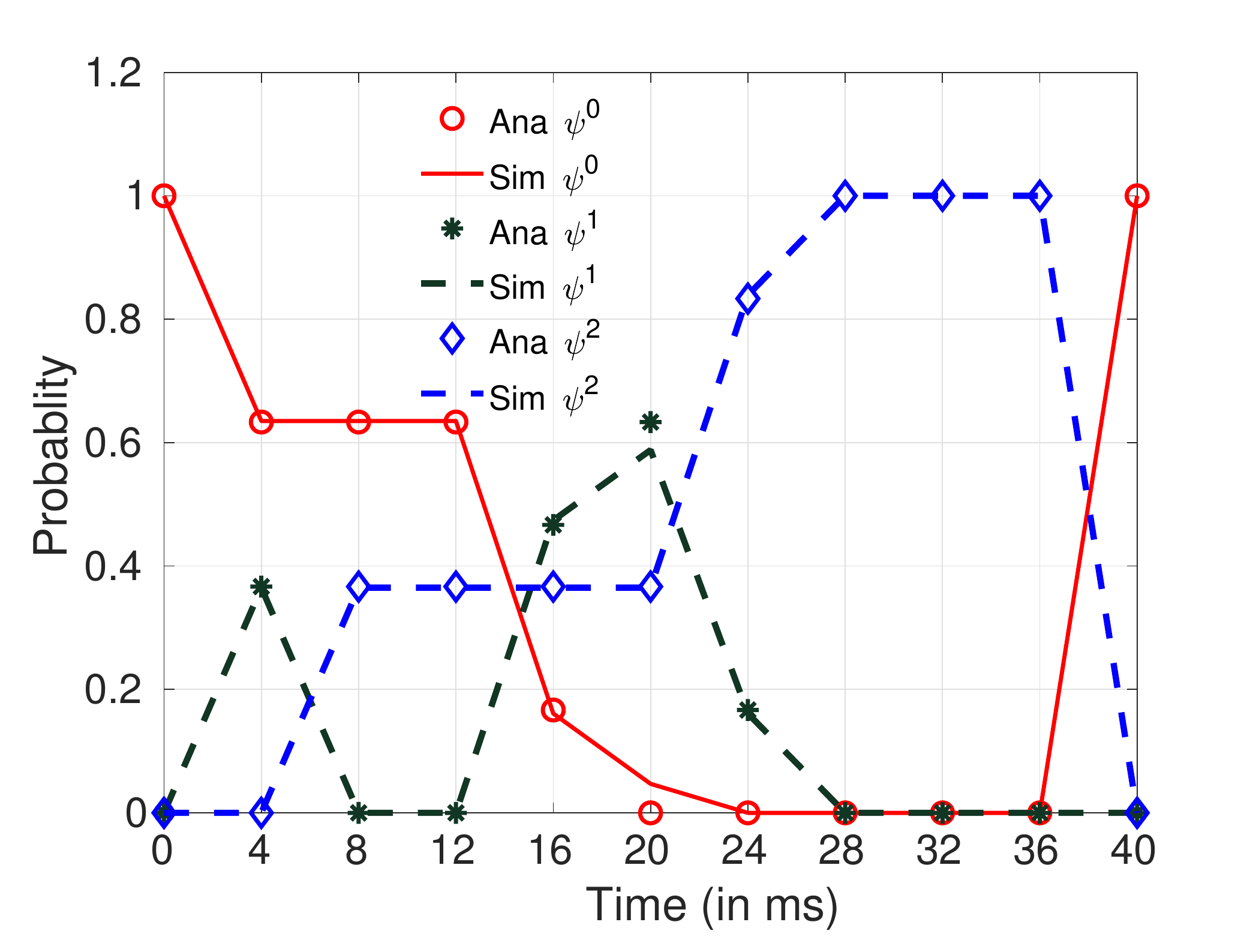}
		\caption{State Probability ($\psi^{\{0,1,2\}}$) of node L4.}
		\label{L4}
	\end{subfigure}
	\begin{subfigure}{0.49\linewidth}
		\includegraphics[width=1.05\linewidth]{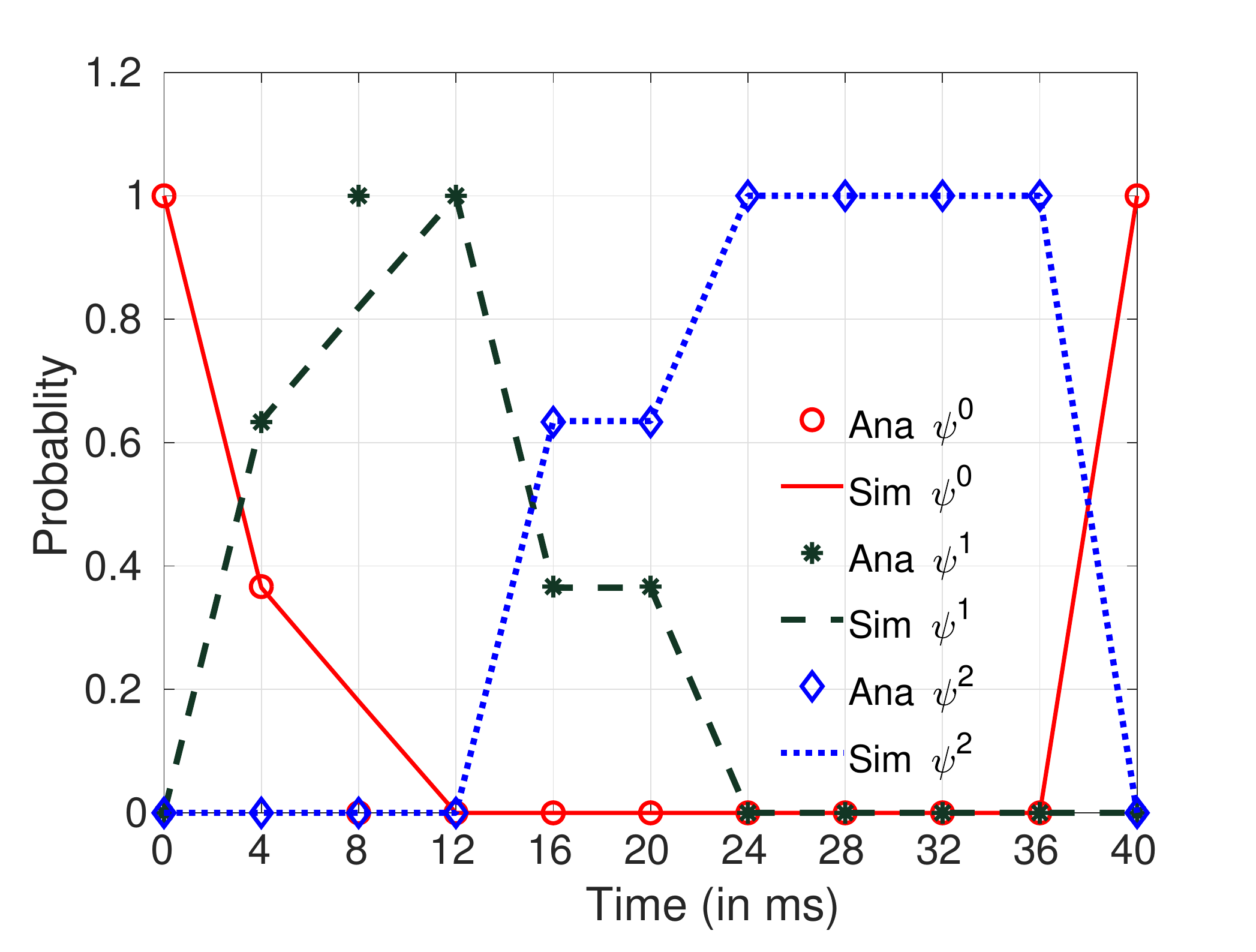}
		\caption{State Probability ($\psi^{\{0,1,2\}}$) of node L5.}
		\label{L5}
	\end{subfigure}
	\caption{Validation of Simulation and Analytical results of LTE-U State Probability for Topology \#2.}
	\label{ProbPred}
\end{figure*}
\subsection{Prediction of LTE-U state in the network}
Eqn.~(\ref{probpred}) in Section~\ref{model} predicts the state of LTE-U at a particular moment of time which can be one among \{0,1,2\}. Our theoretical predictions match with the simulation results. The simulation results are obtained by performing 1000 simulation runs.  Both simulation and analytical results shown in Fig.~\ref{ProbPred} are for Topology~\#2 in Fig.~\ref{scenarios}. In Fig.~\ref{ProbPred}, X-axis and Y-axis indicate time within $T_{frame}$ and probability of LTE-U state, respectively. We observed a close match in predicting the state of \mbox{LTE-U}. Knowing the state of an \mbox{LTE-U} node in future is desirable in designing centralized resource allocation schemes for better spectral utilization. The \mbox{Wi-Fi} nodes that have chances of being victim can be signaled to be muted when an \mbox{LTE-U} node is about to make the transition to prevent channel wastage. We notice that the distribution of state probability of an \mbox{LTE-U} node over time depends on the number of \mbox{LTE-U} nodes around it. Consider $L1$ and  $L5$ nodes of Topology \#2 shown in Fig.~\ref{scenarios}. It is clear that node $L1$ is surrounded by more \mbox{LTE-U} nodes than node $L5$. \mbox{LTE-U} is greedy and starts transmission as soon as it gets access to the channel. This point can be verified from our state probability prediction plots shown in Fig.~\ref{ProbPred}. We can see the probability of node $L5$ to be in a state of active transmission at the beginning of $T_{frame}$ is much higher than that of node $L1$. We can observe $\psi^{1}_{1}$ is spread all across $T_{frame}$ unlike $\psi^{1}_{5}$ which is concentrated mainly at the beginning of $T_{frame}$. This is due to rescheduling behavior of \mbox{LTE-U} node which prevents the channel wastage. \mbox{LTE-U} nodes in such densely concentrated regions will be having a lesser $T_{ON}$ value because of inverse relationship between $T_{ON}$ and number of other nodes on the shared channel. \mbox{Wi-Fi} nodes located in such regions are more likely to suffer from collisions because of frequent transitions among \mbox{LTE-U} nodes. There is a need for better \mbox{LTE-U} mechanism to handle such densely distributed scenarios for achieving better inter-RAT coordination.
\subsection{Coexistence study}
\begin{figure*}[htb!]
	\vspace{-0.3cm}

	\begin{subfigure}{0.5\linewidth}
		\includegraphics[width=0.9\linewidth]{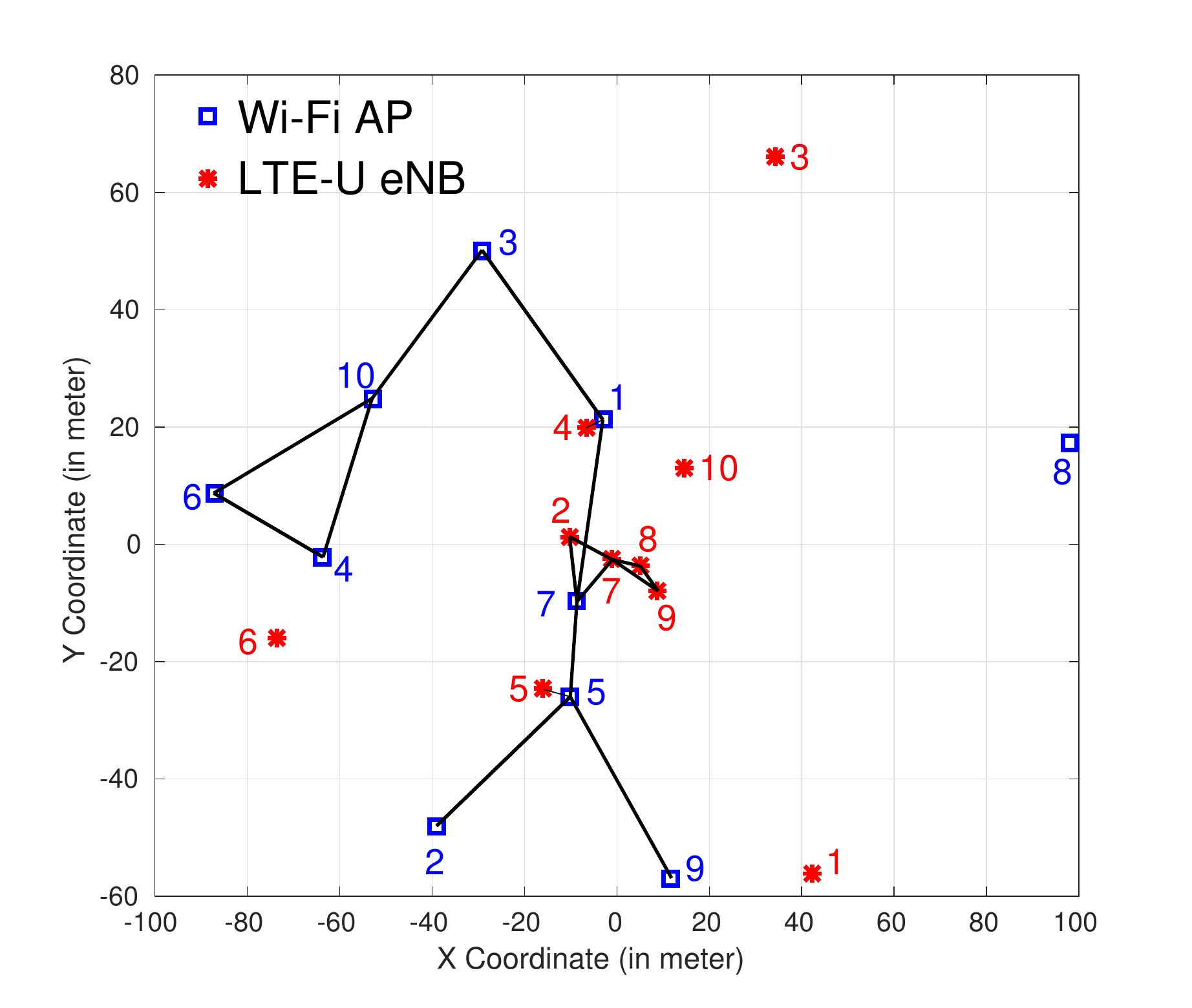}
		\caption{Wi-Fi--LTE-U contention graph.}
		\label{WLG}
	\end{subfigure}
	\begin{subfigure}{0.5\linewidth}

		\includegraphics[width=0.9\linewidth]{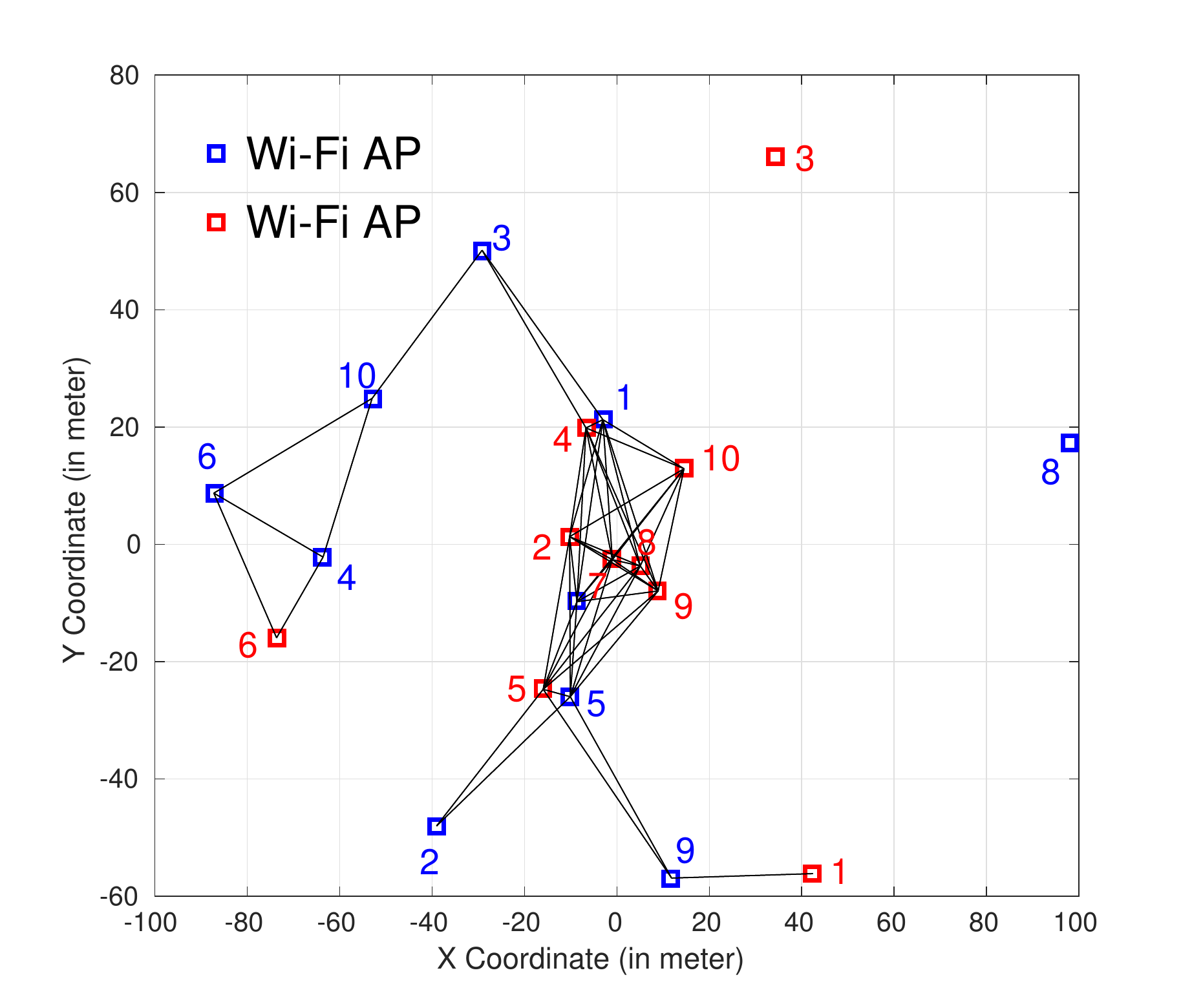}
		\caption{Wi-Fi--Wi-Fi contention graph.}
		\label{WWG}
	\end{subfigure}
	\hspace{-0.3cm}
	\caption{Wi-Fi--LTE-U and Wi-Fi--Wi-Fi contention graphs for one of the randomly generated topologies.}
	\vspace{-0.4cm}
	\label{1000}
\end{figure*}

\begin{table}[htb!]
	\caption{Comparison of WW and WL throughputs (Mbps) in LTE-U and Wi-Fi coexistence study } 
	\centering
	\begin{tabular}{|c|>{\columncolor[gray]{0.8}}c|c|c|}
		\hline
		\\[-1em]
		\bfseries \hspace{-0.25cm} Node Index \hspace{-0.25cm} &\multicolumn{2}{|c|}{\bfseries Fig.~\ref{1000} Topology} \\\hline
		\\[-1em]
		- & WL & WW \\\hline
		\\[-1em]
		W1 & 37.08 & 37.08   \\\hline
		\\[-1em]
		W2 & 37.08 & 74.16 \\\hline
		\\[-1em]
		W3 & 74.16 & 0  \\\hline
		\\[-1em]
		W4 & 49.44 & 0 \\\hline
		\\[-1em]
		W5 & 74.16&  74.16  \\\hline
\\[-1em]		
		W6 & 24.72 & 0   \\\hline
		\\[-1em]
		W7 & 37.08 & 0  \\\hline
		\\[-1em]
		W8 & 18.54 & 0  \\\hline
		\\[-1em]
		W9 & 55.62 & 74.16 \\\hline
		\\[-1em]
		W10 & 37.08&  37.07 \\\hline
	\end{tabular}\label{T4}
\end{table}
After validating the proposed analytical model, we used the model to perform a coexistence study between \mbox{LTE-U} and \mbox{Wi-Fi} nodes. The motivation behind the coexistence study is to check whether \mbox{LTE-U} is fairly coexisting with \mbox{Wi-Fi} or not. According to the 3GPP~\cite{3GPP} definition of fairness, \textit{LTE design in unlicensed bands should be in such a way that it should not impact \mbox{Wi-Fi} more than another \mbox{Wi-Fi} network on the same unlicensed channel.} Thus, the coexistence study is performed by replacing \mbox{LTE-U} nodes with \mbox{Wi-Fi} nodes and studying network behavior in \mbox{Wi-Fi}--\mbox{LTE-U} (WL) and \mbox{Wi-Fi}--\mbox{Wi-Fi} (WW) scenarios. We performed coexistence studies for 20-node over 1000 randomly generated topologies. In WL scenario, 50\% of nodes are Wi-Fi APs (10 nodes) and rest are LTE-U eNBs (10 nodes) whereas for WW scenario the LTE-U eNBs are replaced with Wi-Fi nodes.  One of the topologies out of 1000 topologies is shown in Fig~\ref{1000}. The corresponding results of the coexistence studies are shown with respect to throughputs in Table~\ref{T4}. Table~\ref{T4} shows the throughput performance of \mbox{Wi-Fi} nodes in the presence of \mbox{LTE-U} nodes (\emph{i.e.,} WL scenario) and replacing \mbox{LTE-U} nodes by \mbox{Wi-Fi} nodes (\emph{i.e.,} WW scenario). In Table~\ref{T4}, we can see that the throughput performance of Wi-Fi in WL scenario is better compared to WW scenario. The reason behind throughput improvement of Wi-Fi in WL scenario compared to WW scenario is because of change in the contention graph (shown in Figs.~\ref{WLG} and~\ref{WWG}) and nature of LTE-U protocol to give equal channel share to the neighbouring nodes. Throughput of some of the Wi-Fi nodes is zero because of neighbouring Wi-Fi nodes. For instance, consider W3 and W4 nodes, both the nodes throughput has dropped to zero because of replacing LTE-U with \mbox{Wi-Fi} around it. This can be explained by observing LTE-U protocol. LTE-U protocol is designed to allot dedicated time to \mbox{Wi-Fi} transmissions which is not the case in \mbox{Wi-Fi} networks. W3 and W4 being in the range of many \mbox{Wi-Fi} nodes (once LTE-U nodes are replaced with \mbox{Wi-Fi} nodes), they sense channel to be always busy because of ongoing parallel transmission of \mbox{Wi-Fi} which results in throughput reduction. On the other hand when surrounded by LTE-U nodes, the LTE-U nodes complete their transmissions and leave the channel free for \mbox{Wi-Fi} to transmit. We can observe this point by looking at throughput values in Table~\ref{T4}. So, we can conclude that even though LTE-U and \mbox{Wi-Fi} nodes do not have any coordination to reduce collisions, the fair air time sharing by LTE-U protocol with Wi-Fi can be seen. Thus LTE-U nodes prove to be a better company to \mbox{Wi-Fi} than \mbox{Wi-Fi} nodes.
\par
\begin{figure}[htb!]
	\begin{center}
		\includegraphics[width=0.95\linewidth]{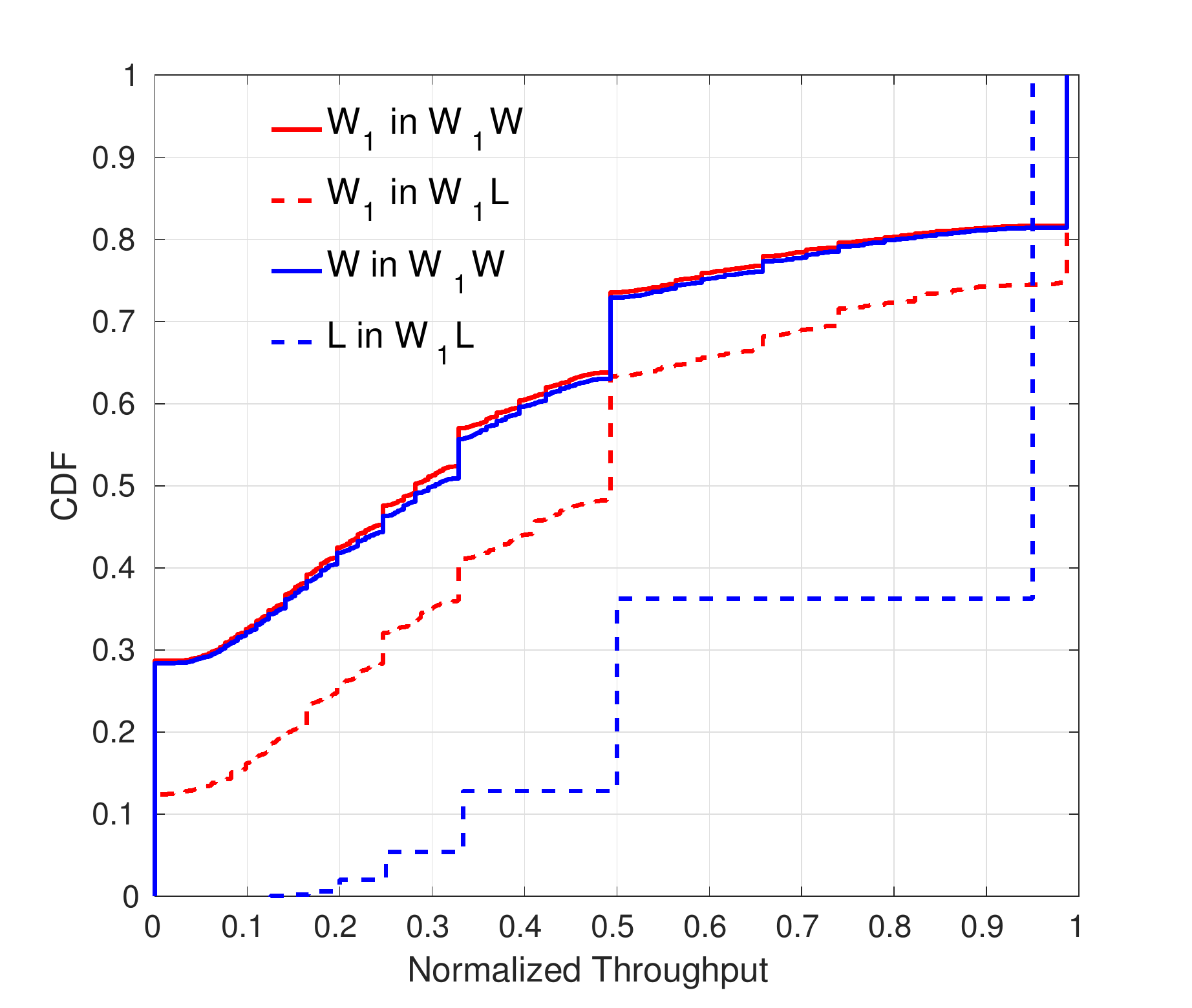}
		\caption{Normalized throughput CDF.}
		\label{CDF}
	\end{center}
	\vspace{-0.7cm}
\end{figure}

 Finally, in Fig.~\ref{CDF}, we have plotted throughput CDF of all the nodes in the network over 1000 randomly generated topologies. The CDF is for 50\% fixed set of Wi-Fi nodes (shown as $W_1$) in WW and WL scenarios. Further, we also plotted CDF for left over Wi-Fi in WW scenario and LTE-U in WL scenario. It is clear from figure that the throughput of Wi-Fi nodes has improved in WL scenario compared to WW scenario. This proves that LTE-U is fair with Wi-Fi in terms of throughput according to 3GPP definition of fairness. In addition, the throughput of left over Wi-Fi in WW scenario is less compared to left over LTE-U in WL scenario. Thus, not only LTE-U is better neighbour to Wi-Fi but also it is better for overall system throughput improvement.
\section{Conclusions}\label{conclusion}
In this paper, we proposed a new analytical model to estimate the throughput of spatially distributed \mbox{LTE-U} and \mbox{Wi-Fi} networks. We used the model to study coexistence of \mbox{LTE-U} and \mbox{Wi-Fi} networks in different scenarios such as all nodes are inside EDT, all nodes are outside EDT, and some are inside EDT and some are outside EDT. The model is based on the probabilistic approach of LTE-U nodes state transitions which effect the behavior of Wi-Fi nodes. When met with proper constraints, the LTE-U nodes can independently transit from one state to other. The behavior of LTE-U node has a great influence on Wi-Fi nodes resulting in throughput variation. Our model defines the relation between LTE-U and Wi-Fi nodes. The model is validated with extensive simulation~studies. Further, we have used the model to analyze the performance of \mbox{Wi-Fi} network in \mbox{LTE-U}--\mbox{Wi-Fi} and \mbox{Wi-Fi}--\mbox{Wi-Fi} networks. From the coexistence study, we found that in spatially distributed scenarios, performance of Wi-Fi is better in LTE-U--Wi-Fi scenario compared to Wi-Fi--Wi-Fi scenario.

\section*{ACKNOWLEDGMENT}
This work was supported by the project “Converged Cloud Communication Technologies”, Meity, Govt. of India.

\bibliographystyle{IEEEtran}
\bibliography{references}
\end{document}